\documentclass[11pt]{article}
\usepackage{iftex}
\RequirePDFTeX
\usepackage[a4paper,margin=2.4cm]{geometry}
\usepackage{newtxtext,newtxmath}
\usepackage[numbers,super,sort&compress]{natbib}
\usepackage[font=small,labelfont=bf,labelsep=period,justification=justified]{caption}
\usepackage{lineno}

\usepackage[utf8]{inputenc}
\usepackage[T1]{fontenc}
\usepackage{graphicx,booktabs,amsmath}
\usepackage{placeins}
\usepackage[hidelinks]{hyperref}
\DeclareUnicodeCharacter{2212}{\ensuremath{-}}
\DeclareUnicodeCharacter{00B5}{\ensuremath{\mu}}
\DeclareUnicodeCharacter{03C7}{\ensuremath{\chi}}

\title{\Large\bfseries Testing correlated spin--orbit multiplets beyond excitation centroids}
\author{Y\"ucel Y\i ld\i r\i m$^{1*}$, Cengiz \c{S}en$^{2}$, \"Ozg\"ur Polat$^{3}$\\[3pt]
\small $^1$Department of Mechanical Engineering, Faculty of Engineering and Natural Sciences,\\
\small Istanbul Bilgi University, Ey\"upsultan 34060, Istanbul, T\"urkiye\\
\small $^2$Department of Physics, Lamar University, Beaumont, TX 77710, USA\\
\small $^3$Department of Physics and Astronomy, California State Polytechnic University,\\
\small 3801 West Temple Ave, Pomona, CA 91768, USA\\
\small $^*$Correspondence: \texttt{yucel.yildirim@bilgi.edu.tr}}
\date{}

\usepackage{microtype}

\hypersetup{colorlinks=true,linkcolor=black,citecolor=black,urlcolor=black}
\hypersetup{pdftitle={Testing correlated spin-orbit multiplets beyond excitation centroids},pdfauthor={Yücel Yıldırım; Cengiz Şen; Özgür Polat},pdfsubject={}}
\begin{document}
\maketitle

\begin{abstract}
\noindent Matching an excitation centroid does not establish the resolved structure of a correlated spin--orbit multiplet. Across eight embedded iridium and osmium compounds, ligand-active XMS-CASPT2 places the K$_2$OsCl$_6$ centroid within 0.5~meV of a RIXS maximum while splitting its nearly degenerate reference triplet by 2.54~meV, and matched OpenMolcas 26.10 controls retain the bare-anion sensitivities. Independent BAGEL calculations isolate contraction dependence in a three-state spin-free model: on the same reference, replacing SS--SR with MS--MR reduces the symmetry-defined E-block splitting from 0.448~meV to $4.67\times10^{-5}$~meV. Native reconstruction, reference symmetry, representation invariance, sensitivity controls and comparison with fitted spectral features or magnetic response provide an ordered validation procedure whose tests constrain different properties of the Hamiltonian.
\end{abstract}
\section*{Introduction}
Multireference perturbation theory is widely used for spin--orbit multiplets of open-shell 4$d$ and 5$d$ compounds. A CASSCF reference captures the static correlation of the metal shell\cite{Roos1980}, second-order perturbation theory approximates external correlation\cite{Andersson1990,Andersson1992}, and a state-interaction step couples the correlated spin-free states through spin--orbit integrals\cite{Malmqvist2002,Hess1996}. Multistate variants let the reference states mix under the correlated Hamiltonian\cite{Finley1998}, and extended multistate formulations address invariance under rotations of the reference space\cite{Shiozaki2011,Granovsky2011}. In practice a calculation also involves an IPEA shift in the zeroth-order Hamiltonian\cite{Ghigo2004,Zobel2017,Nishimoto2023}, an imaginary level shift, a choice of internal contraction\cite{Park2019,Vlaisavljevich2016,Battaglia2020,Park2020} and a selected window of spin-free roots. Each of these can affect quantities much smaller than the excitation energy.

Those small quantities are often the point of the calculation. In $d^4$ and $d^5$ iridates and osmates the splitting of the first excited multiplet by a trigonal or tetragonal field is tens of meV or less, an order of magnitude below the excitation energy, and it controls magnetic anisotropy and the polarization dependence of RIXS\cite{Rossi2017,Kusch2018,Yuan2017,Warzanowski2023}. A common validation step compares energies: a computed centroid is compared with a RIXS maximum or an optical level, but agreement alone does not establish the resolved spectrum. That inference is not safe. A centroid is a trace and is insensitive to how the levels are distributed; a RIXS maximum is not a static level; and multistate perturbation theory can break degeneracies that the symmetry of the problem protects.

This paper turns these concerns into tests and applies them to a common data set: eight embedded clusters spanning $d^4$ and $d^5$ ions, fluoride, chloride, bromide and oxide ligands, analysed with MS- and XMS-CASPT2 at matched geometry, basis and embedding. We separate four kinds of defect that are easily confused: numerical symmetry breaking in the input, representation dependence, sensitivity of the perturbative formulation or its implementation, and errors of state selection. For the implementation question we add an exactly symmetric model, a comparison of two OpenMolcas releases and an independent XMS implementation. The tests are then combined into an ordered protocol in which each step detects failures that the earlier steps cannot.

\section*{Results}
\subsection*{Centroid agreement and the resolved multiplet}
\begin{figure}[!htbp]
\centering
\includegraphics[width=\textwidth]{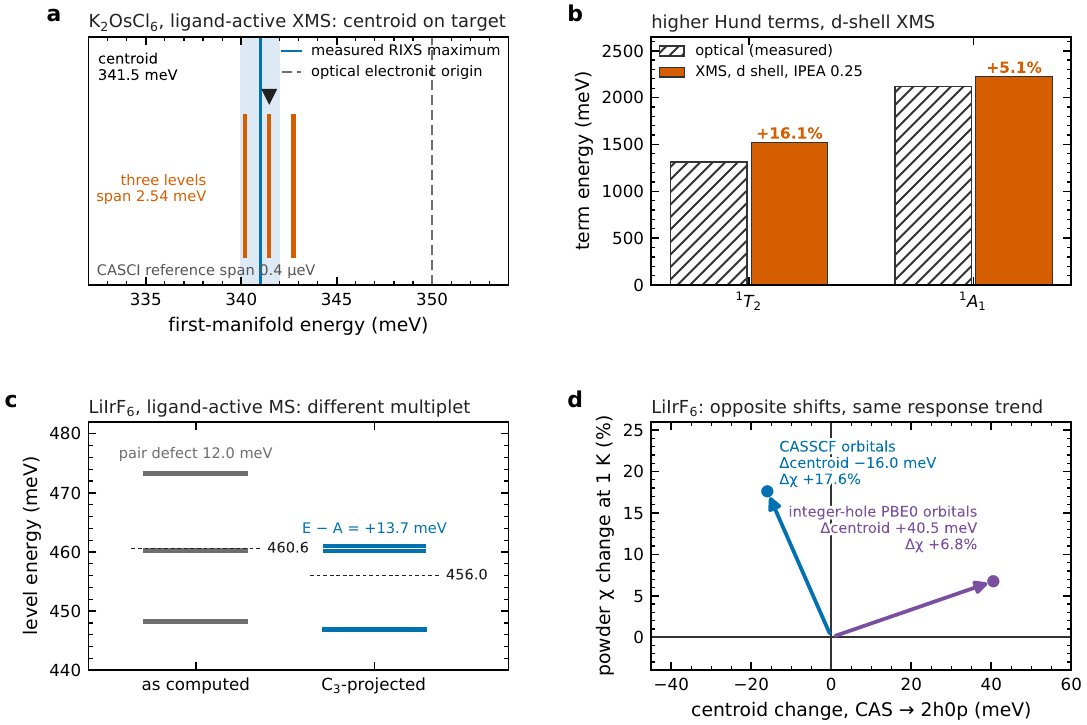}
\caption{\textbf{A correct centroid does not certify the multiplet.} \textbf{a}, K$_2$OsCl$_6$, ligand-active XMS-CASPT2: the first-manifold levels (sticks) and their centroid against the measured RIXS maximum (blue line, band $\pm1$~meV) and the optical electronic origin (dashed). The CASCI reference holds the three levels degenerate. \textbf{b}, Higher singlet terms of K$_2$OsCl$_6$ in the $d$-shell XMS calculation against optical levels. \textbf{c}, LiIrF$_6$, ligand-active MS-CASPT2: levels as computed and after diagnostic $C_3$ projection at zero reference rotation, retaining the same 33/32/15 singlet/triplet/quintet SOC window; dashed lines mark the respective 460.591 and 456.040~meV centroids. The two projected E sticks are displaced by $\pm0.45$~meV solely for visibility; their physical energies coincide. \textbf{d}, LiIrF$_6$: change of the first centroid and of the powder susceptibility at 1~K on opening the 2h0p ligand sector at fixed orbitals, for two orbital-reference protocol families (not a common-limit convergence series). All values are deterministic model results.}
\label{fig:concept}
\end{figure}

K$_2$OsCl$_6$ shows the problem in its simplest form (Fig.~\ref{fig:concept}a). The ligand-active XMS-CASPT2 calculation places the first centroid at 341.5~meV, 0.5~meV from the measured RIXS maximum of 341~$\pm$~1~meV\cite{Warzanowski2023}. The centroid therefore lies within the reported RIXS uncertainty. The three levels of the nominal $J=1$ triplet, however, span 2.54~meV, although the CASCI reference of the same cubic cluster holds them degenerate to 0.4~$\mu$eV. Nothing in the model lifts that degeneracy, so the splitting is produced by the method. The comparison is also less direct than it looks: a RIXS maximum is not a static electronic level, and the optical electronic origin of the same compound lies at 350~meV\cite{Warzanowski2023}, farther from the calculated centroid than the RIXS feature.

Higher multiplets tell a third story (Fig.~\ref{fig:concept}b). In the $d$-shell XMS calculation, whose first excitation lies 6.5\% above the optical origin, the states carrying $^1T_2$ and $^1A_1$ character (mean parent weights 0.73 and 0.87) sit at 1522 and 2224~meV, against optical levels at 1311 and 2117~meV: errors of +16.1\% and +5.1\%. A first excitation close to experiment therefore certifies neither the resolved triplet nor the Hund multiplets above it.

\begin{figure}[!htbp]
\centering
\includegraphics[width=\textwidth]{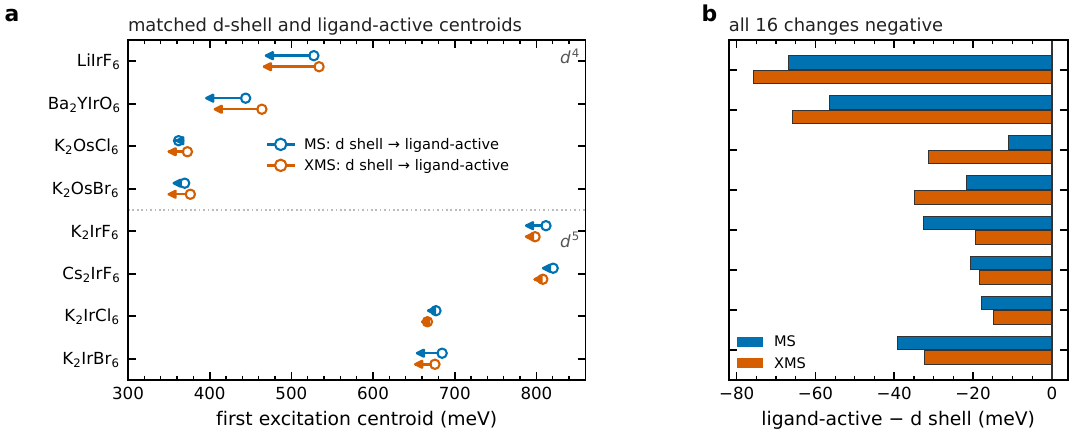}
\caption{\textbf{Ligand-space shifts across eight compounds.} \textbf{a}, First centroids of the $d$-shell (open circles) and ligand-active (arrow heads) calculations, MS (blue) and XMS (orange). \textbf{b}, Ligand-active minus $d$-shell centroid. Geometry, basis and embedding are identical within each compound; MS and XMS are different perturbative variants, not an uncertainty range.}
\label{fig:eight}
\end{figure}

The same tension runs through the eight compounds (Fig.~\ref{fig:eight}). Adding the ligand space at fixed geometry, basis and embedding lowers the first centroid in all 16 MS and XMS comparisons, by 11 to 76~meV. The shifts are systematic in sign but differ by a factor of seven between compounds, so they do not amount to a transferable additive correction. They also describe one protocol. When the same ligand configurations are opened variationally at fixed orbitals, the LiIrF$_6$ centroid moves down on one orbital reference and up on another (Fig.~\ref{fig:concept}d), which shows that the sign of a ``ligand correction'' depends on what else changes with it.

For a spectrally isolated $n$-state manifold with projector $P$, first-order perturbation theory makes the distinction explicit:
\begin{equation}
\delta\overline\Delta=\frac{1}{n}\operatorname{Tr}(P\,\delta H\,P)-\delta E_0.
\label{eq:trace}
\end{equation}
A traceless perturbation within the manifold can split its levels without changing their mean at this order; an excitation centroid also includes the ground-state shift. This trace constraint leaves the traceless part of the Hamiltonian undetermined.

\FloatBarrier
\subsection*{Locating the LiIrF$_6$ pair defect}
\begin{figure}[!htbp]
\centering
\includegraphics[width=\textwidth]{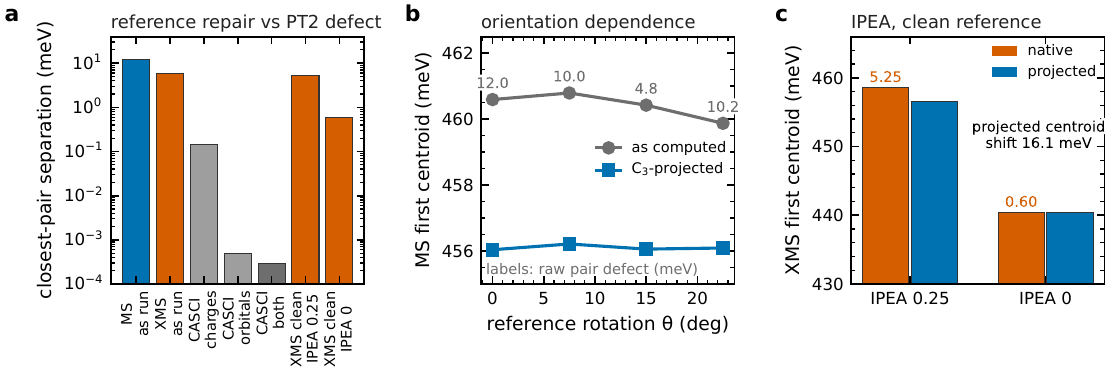}
\caption{\textbf{Locating the LiIrF$_6$ pair defect.} \textbf{a}, Closest-pair separation of the ligand-active MS and XMS spectra as run, of the spin-free CASCI reference after repairing the point charges, the orbital subspaces or both, and of clean-reference XMS at two IPEA values (logarithmic scale; spin-free reference and SOC correlated quantities are distinguished by labels). \textbf{b}, MS first centroid as computed and after $C_3$ projection for four reference rotations; labels give the raw pair separation. \textbf{c}, Clean-reference XMS centroids, native and projected, at IPEA 0.25 and zero; labels give the native pair separation.}
\label{fig:li}
\end{figure}

LiIrF$_6$ has a genuinely trigonal cage, so its E pair is protected by the threefold symmetry together with the even-electron time-reversal structure, and a defect has to be traced before it can be interpreted. The ligand-active MS and XMS centroids are 460.6 and 458.1~meV, with closest-pair separations of 12.0 and 5.7~meV (Fig.~\ref{fig:li}a). For the zero-degree original reference, $C_3$ projection changes the MS centroid from 460.591 to 456.040~meV and changes the multiplet: the as-computed levels place a split pair below a single state, whereas the projected spectrum has A below a degenerate E pair, $13.7$~meV higher (Fig.~\ref{fig:concept}c). The 13.7~meV separation belongs to the projected 456.040~meV operator, not to the raw 460.591~meV spectrum. Both use the same 33-singlet/32-triplet/15-quintet ligand-active SOC window.

The first question is whether the reference already breaks the symmetry. It does, slightly. Symmetrizing the fitted point charges alone leaves a final-CI spin-free pair separation of 0.145~meV, repairing the orbital subspaces alone gives 0.49~$\mu$eV, and both repairs together give 0.29~$\mu$eV. On this clean reference XMS-CASPT2 at IPEA 0.25 still splits the pair by 5.25~meV, with a centroid of 458.6~meV. The original input asymmetry is therefore not sufficient to explain the correlated defect. At IPEA zero the defect falls to 0.60~meV, but the centroid moves by 18.2~meV to 440.4~meV (Fig.~\ref{fig:li}c); a smaller defect does not make this variant more accurate.

The second question is whether the defect depends on a choice that should not matter. Rotating the reference states of the MS calculation by up to $22.5^\circ$, which changes nothing physical, moves the raw pair separation between 4.8 and 12.0~meV (Fig.~\ref{fig:li}b). After projection with a tested threefold representation (angular-momentum fit residual $9.6\times10^{-4}$ for this original reference; Supplementary Note S4) the centroid varies by only 0.18~meV across the four orientations. Projection thus removes a representation dependence that the raw spectrum carries. It does not decide the scalar part of the Hamiltonian: the projected clean-reference centroids at IPEA 0.25 and zero are 456.6 and 440.4~meV, a 16.1~meV difference that symmetry cannot remove.

\FloatBarrier
\subsection*{Exact symmetry, program release and contraction}
\begin{figure}[!htbp]
\centering
\includegraphics[width=\textwidth]{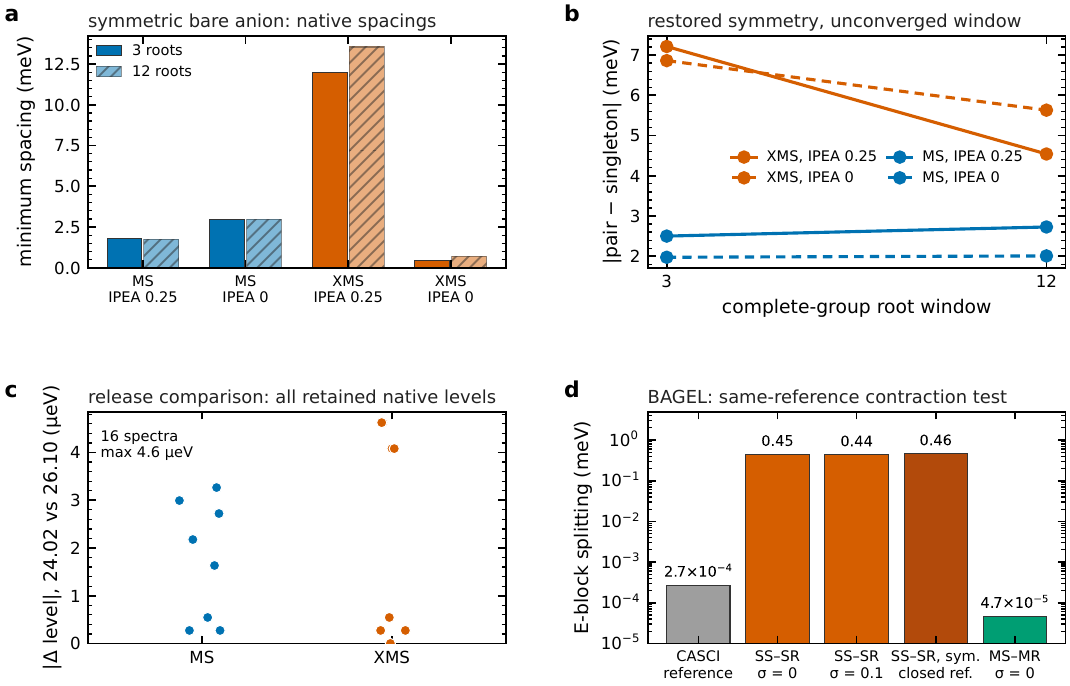}
\caption{\textbf{Exact-symmetry model and implementation controls.} \textbf{a}, Minimum spacing among the three native spin-free levels of the $S_6$ bare anion for MS and XMS at two IPEA values, with 3-root (solid) and 12-root (hatched) complete-group windows; this descriptor is not a symmetry-sector assignment. \textbf{b}, Magnitude of the projected spin-free pair--singleton separation against window size. \textbf{c}, Largest difference over all retained native relative levels between the fresh OpenMolcas 24.02 site baseline and pinned 26.10 for the eight matched cases of each method. \textbf{d}, BAGEL XMS-CASPT2 (three roots, IPEA zero, no frozen orbitals): symmetry-defined E-block splitting of the CASCI reference, of state-specific (SS--SR) calculations at imaginary shifts $\sigma=0$ and 0.1~Hartree, of SS--SR on a symmetry-closed reference, and of the multistate (MS--MR) contraction on the original reference (logarithmic scale).}
\label{fig:impl}
\end{figure}

An $S_6$-symmetric bare [IrF$_6$]$^-$ model separates the reference geometry from the correlated approximation (Fig.~\ref{fig:impl}a). In its three-root spin-free manifold, the minimum spacing among native XMS levels is 11.979~meV at IPEA 0.25 and 0.439~meV at IPEA zero; MS gives 1.776 and 2.978~meV. These are closest-level descriptors, not assignments to an E pair. Projection restores the pair exactly, but it does not fix the window: the projected spin-free pair--singleton separation changes from 7.21 to 4.54~meV when the window grows from 3 to 12 complete-group roots at IPEA 0.25, and from 6.86 to 5.63~meV at IPEA zero (Fig.~\ref{fig:impl}b). MS is window-stable to a few tenths of a meV but depends on the reference orientation. A nine-root window that cuts through a degenerate group raises the representation fit residual from \ensuremath{1.3\times10^{−7}} to 0.38, so the residual detects an invalid window.

These bare-anion findings were obtained with OpenMolcas 24.02. Repeating the eight combinations of root count, IPEA and rotation in the fresh site baseline and pinned 26.10 release reproduces the native relative levels to 3.3~$\mu$eV for MS and 4.6~$\mu$eV for XMS (Fig.~\ref{fig:impl}c). The orientation, IPEA and window sensitivities are therefore retained in both tested builds; the comparison does not isolate source changes from compiler and runtime differences.

An independent spin-free calculation tests the contraction on a common saved reference. With three roots, IPEA zero and no frozen orbitals, the OpenMolcas and BAGEL state-specific calculations give lower-pair spacings of 0.453 and 0.459~meV at zero imaginary shift. Their tightly fitted CASCI gaps still differ by 0.0063~meV, which exceeds the initial 0.001~meV matching target. The comparison is therefore bounded rather than an exact cross-code replica. Inside BAGEL we then change only the first-order space, from single-state single-reference to multistate multireference contraction, with the same saved reference, fitting basis, Fock matrix and XMS rotation. The symmetry-defined E-block splitting falls from 0.448~meV to $\ensuremath{4.7\times10^{−5}}$~meV, and the Frobenius norm of the threefold commutator from 0.885~meV to 0.40~$\mu$eV, below the 0.64~$\mu$eV of the CASCI reference itself (Fig.~\ref{fig:impl}d). Tightening the solver threshold from $10^{-10}$ to $10^{-12}$ leaves the state-specific energies unchanged at the ten-decimal-Hartree output precision. Closing the symmetry of the orbital projectors and recomputing the CI reduces the active-space leakage from \ensuremath{1.3\times10^{−5}} to \ensuremath{1.7\times10^{−9}}, yet the state-specific calculation at 0.1~Hartree shift still has an E-block splitting of 0.464~meV. The same-reference contrast isolates a strong contraction dependence in this three-state model; the small MS--MR residual is not asserted to be identically zero. The formal XMS construction is invariant\cite{Shiozaki2011,Granovsky2011}, and the dependence of practical implementations on contraction has been discussed before\cite{Park2019,Battaglia2020,Vlaisavljevich2016}; what these controls add is a quantified heavy-halide example and a way to separate contraction, reference and observable errors. The bare-anion result does not by itself assign the origin of every embedded SOC defect.

A separate common-reference ORCA comparison tests alternative treatments rather than an independent MS/XMS implementation. From a LiIrF$_6$ CASSCF centroid of 545.344~meV, SC-NEVPT2, QD-NEVPT2 and order-1 DCD-CAS(2) give 566.094, 559.703 and 559.484~meV, with closest-pair spacings of 0.118, 0.560 and 0.000318~meV, respectively\cite{Angeli2001,Angeli2002,Angeli2004,Lang2019,Neese2025}. All 210 CASSCF SOC levels agree between the method preparations. The QD and DCD centroids differ by only 0.219~meV, although their pair defects differ markedly. This supports treating the excitation scale and fine structure as separate tests; it does not establish the accuracy of transformed QD response operators (Supplementary Note S8).

\FloatBarrier
\subsection*{Comparison with measured observables}
\begin{figure}[!htbp]
\centering
\includegraphics[width=\textwidth]{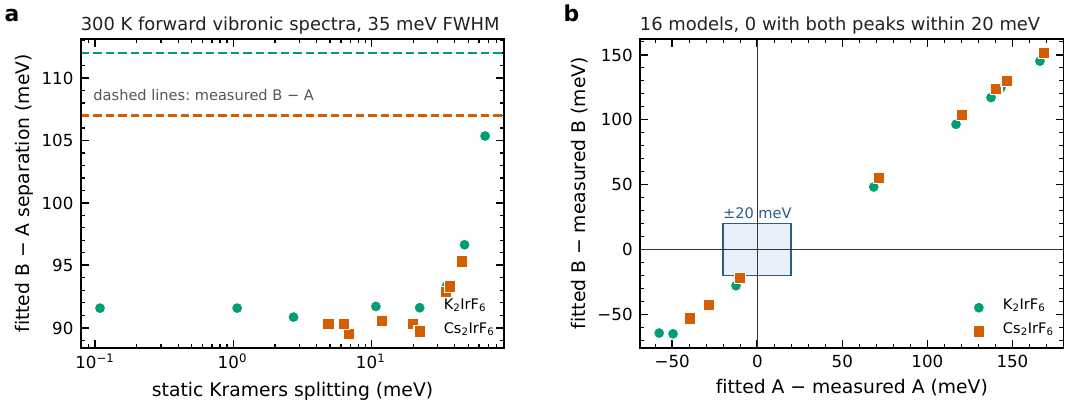}
\caption{\textbf{Fluoride RIXS features compared as the measured observable.} \textbf{a}, Separation of the two fitted features of 300~K forward vibronic spectra (35~meV FWHM) against the static Kramers splitting of each electronic model; dashed lines give the measured separations. \textbf{b}, Fitted minus measured positions of features A and B; the shaded box marks $\pm20$~meV. The vibronic model omits intermediate-state and polarization factors.}
\label{fig:d5}
\end{figure}

Comparison with experiment needs the observable that was measured. For K$_2$IrF$_6$ and Cs$_2$IrF$_6$ the data are two features of 300~K powder RIXS spectra\cite{Rossi2017}, separated by 112 and 107~meV. Propagating each model's static Kramers levels through the $\Gamma_8\otimes(e+t_2)$ vibronic problem and fitting the synthetic spectra with the experimental two-component estimator (Fig.~\ref{fig:d5}) changes the picture considerably. Models with a small static splitting give two features about 90~meV apart, the second being the first $e_g$ sideband rather than a second electronic level. Only models with static splittings of several tens of meV approach the measured separation, and the closest reaches 105.4~meV. None of the 16 models places both fitted centres within 20~meV of experiment. The analysis makes no claim to describe the intermediate-state physics or polarization dependence of the measurement. It shows instead that a static level scheme and a fitted RIXS maximum are not interchangeable, and that a benchmark target derived from one by adding a model-specific offset carries that model into the comparison.

\begin{figure}[!htbp]
\centering
\includegraphics[width=0.78\textwidth]{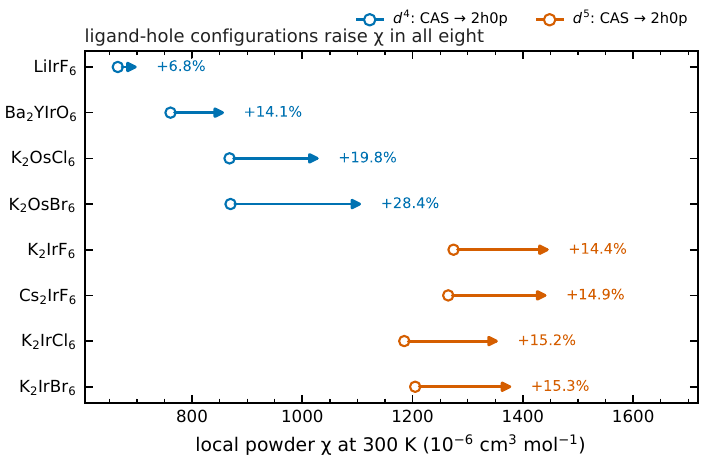}
\caption{\textbf{Ligand-hole configurations raise the single-ion response.} Powder susceptibility at 300~K from complete finite-space CAS (open) and 2h0p (arrow head) response at fixed orbitals for the eight compounds, with the relative change. Single-ion electronic responses without exchange or order.}
\label{fig:fam}
\end{figure}

The magnetic response adds a constraint that energies alone do not supply (Fig.~\ref{fig:fam}). Opening the 2h0p ligand sector at fixed orbitals raises the 300~K single-ion susceptibility in all eight compounds, by 6.8--28.4\% for the $d^4$ ions and 14.4--15.3\% for the $d^5$ ions. In LiIrF$_6$ the same configuration extension lowers the centroid by 16.0~meV on CASSCF orbitals and raises it by 40.5~meV on integer-hole PBE0 orbitals, yet the 1~K susceptibility rises by 17.6\% and 6.8\% (Fig.~\ref{fig:concept}d). The response depends on transition matrix elements and on states well above the first manifold, so it is not a function of the lowest gap. These response trajectories compare an earlier 132-electron CASSCF-derived model with a 144-electron PBE0-derived model, so their cross-family difference includes basis and boundary changes. A companion study uses matched basis and boundary controls to examine the fine structure under an orbital switch\cite{CompanionMechanism}. These are single-ion responses; K$_2$IrCl$_6$ and K$_2$IrBr$_6$ order antiferromagnetically at low temperature\cite{Khan2019,ReigiPlessis2020,Wang2025,Khan2021}, K$_2$IrF$_6$ shows hyperfine evidence of order at very low temperature\cite{Potzel1977}, and the magnetism of Ba$_2$YIrO$_6$ is sample dependent\cite{Dey2016,Terzic2017}.

The internal partition of the orbitals matters even when their union changes little. In a separate matched-basis Li orbital switch, principal angles from the singular values of $C_A^\dagger S C_B$ reach only $2.406^\circ$ for the ten-orbital metal-plus-ligand union, but $8.131^\circ$ and $7.782^\circ$ for its five-dimensional metal and ligand subspaces. The CAS-to-2h0p centroid increment changes from $-16.141$ to $+40.508$~meV. A restricted configuration rule therefore acts on different many-electron spaces after the orbital replacement. These matched-basis results identify the partition change; the response trajectories in Fig.~\ref{fig:concept}d are separate protocol families and must not be combined into one convergence series (Supplementary Note S7).

The full temperature curves, finite-field controls and decomposition into the first trio and higher virtual states are given in Supplementary Note S7.

At 300~K, residuals relative to the published extracted Van Vleck components change from $-9.58\%$ to $+8.35\%$ for K$_2$OsCl$_6$ and from $-14.72\%$ to $+9.52\%$ for K$_2$OsBr$_6$ on opening the ligand sector; the Ba$_2$YIrO$_6$ residual instead increases from $+1.33\%$ to $+15.65\%$ (Supplementary Table S7e). These sample-dependent extracted components have no quoted extraction uncertainties, so the smaller absolute osmate residuals and the larger oxide residual are descriptive comparisons, rather than a uniform accuracy gain\cite{Warzanowski2023,Dey2016}.

\FloatBarrier
\subsection*{An ordered validation protocol}
\begin{figure}[!htbp]
\centering
\includegraphics[width=\textwidth]{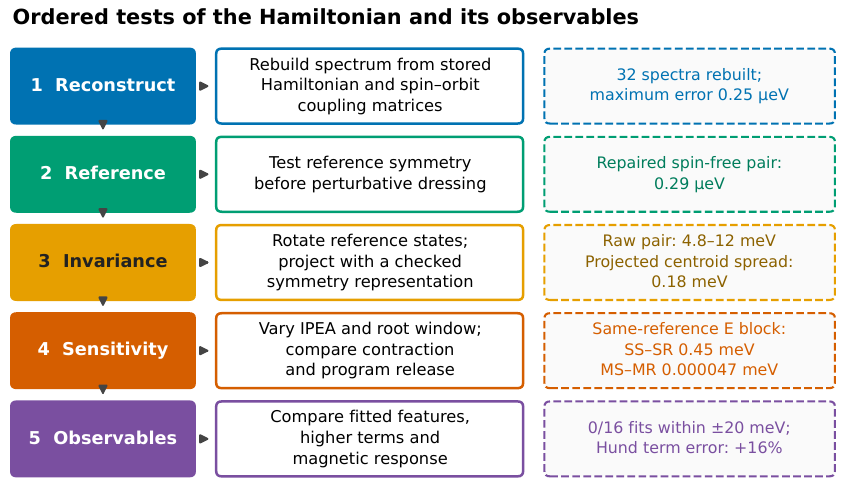}
\caption{\textbf{An ordered validation sequence for correlated spin--orbit multiplets.} Each test isolates a class of defect that the earlier ones cannot detect. Dashed boxes give what the test found in this work.}
\label{fig:protocol}
\end{figure}

The failures above are of different kinds, and each is caught by a different test (Fig.~\ref{fig:protocol} and Table~\ref{tab:classes}). Native reconstruction confirms that the stored matrices describe the computed spectrum: all 32 baseline spectra are rebuilt to 0.25~$\mu$eV. That says nothing about symmetry. A test of the reference before dressing separates numerical symmetry breaking in the input from defects introduced by perturbation theory. Rotating reference states probes invariance under a change that should not matter, and a validated projection then removes the noninvariant part. Varying IPEA, root window, contraction and program release separates sensitivities of the formulation from those of an implementation and from errors of state selection. The last step compares the calculated quantity with the measured observable. Its agreement and uncertainty must be assessed separately from the numerical and symmetry checks.

A failed reconstruction stops interpretation of the parsed operator. A failed reference-symmetry or representation test prevents assigning projected levels as a physical prediction; the enlarged Li window is retained only as a diagnostic. Once those checks pass, a controlled change of the suspected approximation can support causal attribution within that reference. Comparison with spectral or magnetic data is a separate step and uses the stated forward model and transition operators.

\begin{table}[t]
\centering
\caption{\textbf{Classes of defect and the test that isolates each.} Examples are from this work.}
\label{tab:classes}
\small
\begin{tabular}{p{0.16\linewidth}p{0.20\linewidth}p{0.29\linewidth}p{0.24\linewidth}}
\toprule
class & test & example & does not establish\\
\midrule
assembly or parsing & native reconstruction & 32 spectra to 0.25~$\mu$eV & any physical property\\
numerical symmetry breaking in the input & reference commutator and pair & LiIrF$_6$ CASCI pair 0.145~meV $\rightarrow$ 0.29~$\mu$eV & absence of a correlated defect\\
representation dependence & reference rotation; tested projection & raw pair 4.8--12.0~meV; projected centroid spread 0.18~meV & correctness of the symmetric part\\
implementation (contraction) & same reference, SS--SR vs MS--MR & E-block 0.448 $\rightarrow$ $\ensuremath{4.7\times10^{−5}}$~meV & a universal limitation of XMS\\
formulation sensitivity & IPEA; release cross-check & clean-reference projected centroid shift 16.1~meV & which variant is accurate\\
state selection & complete-group windows; fit residual & projected splitting 7.2 $\rightarrow$ 4.5~meV; residual 0.38 & energy convergence\\
observable mismatch & forward line shape; higher terms; response & 0/16 fluoride fits; $^1T_2$ +16\% & accuracy of other observables\\
\bottomrule
\end{tabular}
\end{table}

\FloatBarrier
\section*{Discussion}
The tests distinguish a plausible excitation scale from a validated resolved Hamiltonian. Projection with an inferred, checked representation removes a noninvariant component, but it cannot determine the symmetry-allowed part. In the bare-anion model, projection leaves appreciable state-window sensitivity; in embedded LiIrF$_6$, it leaves a 16.1~meV IPEA dependence. The raw spectrum and its projected diagnostic must therefore be reported together.

Attribution requires controls that change one consequential choice at a time. The two OpenMolcas builds retain the main sensitivities, while the same-reference BAGEL SS--SR/MS--MR comparison isolates contraction dependence in a separate three-root spin-free model. The finite cross-code reference mismatch, fitting-basis error and reference-closure residuals bound that conclusion. These results do not establish universal failure of XMS-CASPT2 or certify every embedded state window. Contraction and IPEA invariance limitations were already known\cite{Battaglia2020,Vlaisavljevich2016,Nishimoto2023}; the present contribution is their quantified consequence for heavy-halide multiplets and an ordered set of tests that separates their effects.

Park compared SS--SR and MS--MR zeroth-order constructions in MS-CASPT2 and their consequences near conical intersections\cite{Park2019}. RMS- and XDW-CASPT2 address related state-mixing and invariance questions through rotated references and state-dependent Fock operators\cite{Battaglia2020,BattagliaLindh2021}. In the symmetry-aware XDW formulation, Hamiltonian couplings enter the weights so that states of different symmetry are not spuriously averaged\cite{BattagliaLindh2021}. Those changes differ from replacing the contracted first-order span on a fixed reference. RMS/XDW have not been tested on the operators examined here, and cannot be identified as cures for their observed defects. The family results also retain their finite-basis scope: the central units use ANO-RCC-VDZP, except for the stated VTZP Ir/O treatment in Ba$_2$YIrO$_6$. The controlled contraction comparison does not establish basis-converged defect magnitudes for the embedded SOC problem.

Observable comparisons add information that symmetry cannot supply. The fluoride calculation compares fitted spectral features after a specified approximate vibronic model, and the finite-space susceptibility depends on magnetic matrix elements and higher states. Energy-only parameter fitting leaves corresponding unconstrained directions: in an explicitly imperfect full-shell Li mapping, compensating Coulomb and SOC changes keep two centroids within 0.041~meV while changing a fine-structure descriptor by about 4~meV. Supplementary Note S8 retains the mapping residual, parameter convention and the separate 488-state downfolding check. Neither is a determination of material parameters from centroid agreement.

For LiIrF$_6$ we do not use a projected perturbative energy as a prediction. The companion study obtains its trigonal fine structure from complete variational configuration spaces and independent DCD-CAS(2) controls\cite{CompanionMechanism}. The two papers address complementary questions: the physical origin of that fine structure, and the tests needed before a correlated multiplet can support such an interpretation.

A matching excitation centroid does not determine degeneracies, level order, higher multiplets or magnetic response. In the eight compounds studied here, centroid agreement coexists with symmetry defects and with sensitivity to orbital reference, IPEA and the retained root window. The matched OpenMolcas 26.10 controls retain the bare-anion findings, while the independent same-reference BAGEL test reduces a 0.448~meV E-block splitting to $4.67\times10^{-5}$~meV by changing the first-order contraction. Native reconstruction, reference symmetry, representation invariance, controlled sensitivity tests and comparison with the measured observable provide distinct evidence. Reporting them separately makes clear which predictions the calculation supports.

\section*{Methods}
\subsection*{Compounds, embedding and active spaces}
The embedded clusters are LiIrF$_6$, Ba$_2$YIrO$_6$, K$_2$OsCl$_6$ and K$_2$OsBr$_6$ ($d^4$), and K$_2$IrF$_6$, Cs$_2$IrF$_6$, K$_2$IrCl$_6$ and K$_2$IrBr$_6$ ($d^5$). Each central $MX_6$ unit is all-electron with ANO-RCC-VDZP basis sets, except for the VTZP Ir/O basis in Ba$_2$YIrO$_6$\cite{Roos2004,Roos2005} and the second-order Douglas--Kroll--Hess Hamiltonian\cite{Hess1986}; neighbouring cations carry embedding \textit{ab initio} model potentials\cite{Huzinaga1987,BarandiaranSeijo1988} and a fitted point-charge array reproduces the Ewald field. For Ba$_2$YIrO$_6$ the Y and Ba sites use model potentials derived for Cs$_2$NaYF$_6$ and BaF$_2$ hosts. Metal-shell references are CAS(4,5) or CAS(5,5); adding five initially occupied ligand combinations gives fourteen or fifteen electrons in ten orbitals. The complete metal-shell root counts are 50/45/5 singlets/triplets/quintets for $d^4$ and 75/24/1 doublets/quartets/sextets for $d^5$. Two active spaces are compared within each compound at identical geometry, basis and embedding: the metal $d$ shell alone and the $d$ shell plus five occupied ligand combinations selected by a fixed projector in the atomic-orbital metric.

\subsection*{Multistate perturbation theory and spin--orbit coupling}
Spin-free reference states from state-averaged CASSCF\cite{Roos1980} are dressed by MS-CASPT2\cite{Finley1998} or XMS-CASPT2\cite{Shiozaki2011,Granovsky2011} in OpenMolcas 24.02 (refs.~\citenum{FdezGalvan2019,LiManni2023}), with IPEA shift 0.25 Hartree\cite{Ghigo2004} and imaginary level shift 0.1 Hartree unless stated, and coupled through atomic mean-field spin--orbit integrals\cite{Hess1996} in a state-interaction step\cite{Malmqvist2002}. Determinant-based restricted spaces use explicit occupation-sector rules\cite{Olsen1988}; the IPEA zeroth-order Hamiltonian can depend on rotations among degenerate active orbitals\cite{Safari2024}. For a given scalar reference the spin--orbit problem is
\begin{equation}
\mathcal H=\bigoplus_S\left(H_S^{\mathrm{PT2}}\otimes\mathbf{1}_{2S+1}\right)+H_\mathrm{SO},
\label{eq:hso}
\end{equation}
with $H_S^{\mathrm{PT2}}$ the correlated effective Hamiltonian of spin block $S$. The first centroid is the mean of the lowest three excited spin--orbit levels for $d^4$ and of the four excited components for $d^5$, relative to the ground state.

\subsection*{Native reconstruction}
Every printed effective Hamiltonian, XMS reference rotation and spin--orbit matrix is transformed into the CASCI reference gauge, the printed energy offsets are restored, and Eq.~(\ref{eq:hso}) is diagonalized again. Rounded XMS transformations are replaced by their nearest orthogonal polar factors after separating the common scalar energy. Absolute and relative energies are checked separately, because an error in a common offset cancels from every excitation energy.

\subsection*{Symmetry representation and projection}
In the CASCI gauge we construct a unitary representation $U$ of the threefold rotation, both from the transformation of the angular-momentum and spin--orbit integrals and from the structure of the reference energy groups, and define
\begin{equation}
H_\mathrm{sym}=\frac13\sum_{k=0}^{2}U^kH\,U^{-k},\qquad U^\dagger U=\mathbf{1},\quad U^3=\mathbf{1}.
\label{eq:hsym}
\end{equation}
$U$ is accepted only if it is unitary, closes, commutes with the reference Hamiltonian, transforms the vector operators covariantly and leaves the spin--orbit matrix invariant; dipole operators withheld from the fit provide an additional covariance test. In the embedded controls this is an inferred operator representation: the angular-momentum and atomic mean-field spin--orbit checks are partly correlated, and no direct many-electron rotation witness is claimed for that cluster. The direct AO/CI rotation witness below applies to the distinct bare-anion model. $C_3$ has one-dimensional complex irreducible representations, so a degenerate E pair in the even-electron problem also needs the time-reversal structure that pairs conjugate characters. Retaining whole energy groups in a state window is a practical safeguard against cutting through such a pair; the representation test, not the energy gap, decides whether the window is invariant. Equation~(\ref{eq:hsym}) removes the part of $H$ that is incompatible with the chosen symmetry. It preserves the trace but not, in general, the excitation centroid, and it cannot tell which symmetric part is correct.

\subsection*{Exact-symmetry model, release comparison and independent implementation}
The minimal model is the bare [IrF$_6$]$^-$ anion at an exactly $S_6$-symmetric geometry, in CAS(4,5) with 36 frozen orbitals, three or twelve triplet roots (whole degenerate groups) and reference rotations of 0 and $15^\circ$. The exact geometry protects an E pair, but finite reference-subspace closure must still be checked. A minimum spacing among sorted native levels is only a descriptor; assigning a symmetry-sector defect requires the representation. The same eight combinations of root count, IPEA and rotation were run in the fresh OpenMolcas 24.02+440-g175b3a12d site build and in the public 26.10 release, compiled from a pinned source commit after five CTest checks passed. This site baseline is distinct from the archived eight-compound 24.02 campaign; compiler and runtime differences remain part of the release comparison. For an independent implementation we used BAGEL 1.2.2 (ref.~\citenum{Shiozaki2018}), whose fully internally contracted XMS-CASPT2 offers a single-state, single-reference (SS--SR) or a multistate, multireference (MS--MR) first-order space\cite{Vlaisavljevich2016,MacLeod2015}. These controls use three triplet roots, IPEA zero and no frozen orbitals, because the two programs choose different frozen-orbital projectors after internal rotations. Within BAGEL the SS--SR and MS--MR runs share the saved reference, fitting basis, state-averaged Fock matrix and XMS rotation; only the contraction differs. Before metric dependencies are removed and the interacting basis is orthogonalized, the two first-order spaces are
\begin{equation}
\begin{split}
\mathcal V_I^{\mathrm{SS\text{-}SR}}&=\operatorname{span}_{\mu}\{\hat E_\mu|\widetilde\Phi_I\rangle\},\\
\mathcal V_I^{\mathrm{MS\text{-}MR}}&=\operatorname{span}_{\mu,J}\{\hat E_\mu|\widetilde\Phi_J\rangle\}.
\end{split}
\label{eq:contraction}
\end{equation}
Both controls retain the same three reference roots. Their symmetry-defined E-block splitting is evaluated from the effective Hamiltonian using the direct AO/CI representation. Fixed closed, active and virtual subspaces are retained through a supported BAGEL input path that stops before interspace orbital updates, rotates within blocks, and recomputes CI and density matrices. This is a fixed-subspace comparison, not a claim of orbital stationarity. The 113-function ANO-RCC orbital basis and point-nucleus DKH2 Hamiltonian are fixed. Both BAGEL contractions use an orthogonal interacting basis, overlap cutoff $10^{-9}$ and residual threshold $10^{-10}$; a $10^{-12}$ control tests solver sensitivity. The direct spatial rotation is reconstructed from actual orbitals and all 50 printed CI determinants, including the closed-shell determinant factor. An enlarged all-primitive-product-range fitting basis and separate tighter-RICD OpenMolcas references test the integral approximation. Supplementary Note S5 gives the numerical fitting convergence, closure residuals and unchanged input records.

\subsection*{Forward vibronic spectra of the $d^5$ fluorides}
For K$_2$IrF$_6$ and Cs$_2$IrF$_6$ the measured quantities are two features of 300~K powder RIXS spectra at 35~meV resolution\cite{Rossi2017}. We propagate the static Kramers doublets of each electronic model through
\begin{equation}
\begin{split}
H_\mathrm{vib}={}&H_\mathrm{stat}+\sum_{\nu=1}^{5}\hbar\omega_\nu b_\nu^\dagger b_\nu\\
&+g_e(q_1 E_\theta+q_2 E_\epsilon)\\
&+g_t(q_3 T_\xi+q_4 T_\eta+q_5 T_\zeta),
\end{split}
\label{eq:hvib}
\end{equation}
with $q_\nu=b_\nu+b_\nu^\dagger$, two $e_g$ and three $t_{2g}$ coordinates, and coupling matrices from symmetric displacements of the embedded cages\cite{Iwahara2023}. The common harmonic zero-point constant is omitted. Couplings computed for the CAS and 2h0p references are transferred to the other electronic models of the same space. Equal weights of the four $\Gamma_8$ components, thermal populations at 300~K and 35~meV FWHM Gaussian broadening define the envelope. The spectra are then fitted, like the experiment, with two Pearson~VII components $a_j[1+((E-E_j)/w_j)^2(2^{1/m}-1)]^{-m}$ sharing one exponent $m$, where $w_j$ is the half width at half maximum.

\subsection*{Fixed-orbital magnetic response}
For each compound the Hamiltonian and magnetic operators are constructed in complete CAS and 2h0p configuration spaces at fixed orbitals. The 2h0p restriction allows up to two holes in five initially occupied ligand orbitals and no occupation of the selected external radial group; its full dimensions are 12,180 for $d^4$ and 7,752 for $d^5$. The zero-field susceptibility is the Kubo response of the full finite space\cite{Kubo1957}, with $\boldsymbol\mu=-\mu_\mathrm B(\mathbf L+g_e\mathbf S)$, analytic degenerate limits and Sternheimer completion of the higher virtual-state contribution. Finite-field checks use $H(\mathbf B)=H(0)-\boldsymbol\mu\cdot\mathbf B$. These are single-ion electronic responses without exchange, order or core diamagnetism.

\section*{Supplementary material}
Cluster definitions and electron counts; native reconstruction and all 32 first-manifold spectra; reference, representation, state-window, release and contraction controls; bare-anion Cartesian coordinates; vibronic Hamiltonian, coupling parameters and fitted spectral features; complete finite-space response, finite-field and temperature controls, matched orbital switch, ORCA comparison, downfolding and parameter-identifiability checks (PDF). Source records, native matrices and replay scripts are supplied separately (ZIP).

\section*{Acknowledgments}
The principal electronic-structure calculations reported in this paper were fully performed using the EuroHPC Joint Undertaking (EuroHPC JU) supercomputer MareNostrum 5, hosted by the Barcelona Supercomputing Center (BSC). Access to MareNostrum 5 was provided through a national access call coordinated by the Scientific and Technological Research Council of Turkey (T\"UB\.{I}TAK). We gratefully acknowledge BSC, T\"UB\.{I}TAK, and the EuroHPC JU for providing access to these resources and supporting this research.

\section*{Competing interests}
The authors declare no competing interests.

\section*{Data availability}
Numerical source tables, native input decks, effective Hamiltonians, orbital and operator records, and the release and contraction controls are deposited in the separate Paper 2 Zenodo archive, version 1.0.0, DOI: \href{https://doi.org/10.5281/zenodo.23917145}{10.5281/zenodo.23917145}. Public access begins when the deposit is published.

\section*{Code availability}
Matrix-reconstruction, forward-spectrum, magnetic-response and figure-generation scripts, with their numerical inputs and replay instructions, are included in the same Zenodo archive, DOI: \href{https://doi.org/10.5281/zenodo.23917145}{10.5281/zenodo.23917145}. The archive distinguishes checks of stored records from fresh numerical solves and electronic-structure calculations. Electronic-structure programs and basis libraries remain separate dependencies under their respective licences. Public access begins when the deposit is published.

\section*{Tool Use}
OpenMolcas was used for CASSCF, MS/XMS-CASPT2 and spin--orbit calculations, including comparisons between software releases. BAGEL was used for independent XMS-CASPT2 contraction comparisons, and ORCA for NEVPT2 and DCD-CAS(2) controls. Analysis scripts were used for Hamiltonian reconstruction, symmetry checks, spectral simulations, magnetic response and plotting. Computational details and software references are provided in Methods and the Supplementary Information.

During the preparation of this manuscript, the authors used generative artificial intelligence assistants for language editing and text organization. No text, data, figures, or scientific interpretations were generated autonomously. Numerical results were obtained using the computational methods described in this work. The authors performed all calculations, verified every number, figure, and reference against primary sources, and take full responsibility for the content of the published article.

\bibliographystyle{naturemag}
\bibliography{references}

@article{Dey2016,
  author = {Dey, T. and Maljuk, A. and Efremov, D. V. and Kataeva, O. and Gass, S. and Blum, C. G. F. and Steckel, F. and Gruner, D. and Ritschel, T. and Wolter, A. U. B. and Geck, J. and Hess, C. and Koepernik, K. and van den Brink, J. and Wurmehl, S. and B{\"u}chner, B.},
  title = {{Ba$_2$YIrO$_6$: A cubic double perovskite material with Ir$^{5+}$ ions}},
  journal = {Physical Review B},
  year = {2016},
  volume = {93},
  pages = {014434},
  doi = {10.1103/PhysRevB.93.014434},
  number = {1},
}

@article{Terzic2017,
  author = {Terzic, J. and Zheng, H. and Ye, Feng and Zhao, H. D. and Schlottmann, P. and De Long, L. E. and Yuan, S. J. and Cao, G.},
  title = {{Evidence for a low-temperature magnetic ground state in double-perovskite iridates with Ir$^{5+}$ ($5d^4$) ions}},
  journal = {Physical Review B},
  year = {2017},
  volume = {96},
  pages = {064436},
  doi = {10.1103/PhysRevB.96.064436},
  number = {6},
}

@article{Kusch2018,
  author = {Kusch, M. and Katukuri, V. M. and Bogdanov, N. A. and B{\"u}chner, B. and Dey, T. and Efremov, D. V. and Hamann-Borrero, J. E. and Kim, B. H. and Krisch, M. and Maljuk, A. and Sala, M. Moretti and Wurmehl, S. and Aslan-Cansever, G. and Sturza, M. and Hozoi, L. and van den Brink, J. and Geck, J.},
  title = {{Observation of heavy spin-orbit excitons propagating in a nonmagnetic background: The case of (Ba,Sr)$_2$YIrO$_6$}},
  journal = {Physical Review B},
  year = {2018},
  volume = {97},
  pages = {064421},
  doi = {10.1103/PhysRevB.97.064421},
  number = {6},
}

@article{Yuan2017,
  author = {Yuan, Bo and Clancy, J. P. and Cook, A. M. and Thompson, C. M. and Greedan, J. and Cao, G. and Jeon, B. C. and Noh, T. W. and Upton, M. H. and Casa, D. and Gog, T. and Paramekanti, A. and Kim, Young-June},
  title = {{Determination of Hund's coupling in $5d$ oxides using resonant inelastic x-ray scattering}},
  journal = {Physical Review B},
  year = {2017},
  volume = {95},
  pages = {235114},
  doi = {10.1103/PhysRevB.95.235114},
  number = {23},
}

@article{Rossi2017,
  author = {Rossi, M. and Retegan, M. and Giacobbe, C. and Fumagalli, R. and Efimenko, A. and Kulka, T. and Wohlfeld, K. and Gubanov, A. I. and Moretti Sala, M.},
  title = {{Possibility to realize spin-orbit-induced correlated physics in iridium fluorides}},
  journal = {Physical Review B},
  year = {2017},
  volume = {95},
  pages = {235161},
  doi = {10.1103/PhysRevB.95.235161},
  number = {23},
}

@article{Iwahara2023,
  author = {Iwahara, Naoya and Furukawa, Wataru},
  title = {{Vibronic effect on resonant inelastic x-ray scattering in cubic iridium hexahalides}},
  journal = {Physical Review B},
  year = {2023},
  volume = {108},
  pages = {075136},
  doi = {10.1103/PhysRevB.108.075136},
  number = {7},
}

@article{Warzanowski2023,
  author = {Warzanowski, P. and Magnaterra, M. and Stein, P. and Schlicht, G. and Faure, Q. and Sahle, Ch. J. and Lorenz, T. and Becker, P. and Bohat{\'y}, L. and Moretti Sala, M. and Monaco, G. and van Loosdrecht, P. H. M. and Gr{\"u}ninger, M.},
  title = {{Electronic excitations in $5d^4$ $J=0$ Os$^{4+}$ halides studied by resonant inelastic x-ray scattering and optical spectroscopy}},
  journal = {Physical Review B},
  year = {2023},
  volume = {108},
  pages = {125120},
  doi = {10.1103/PhysRevB.108.125120},
  number = {12},
}

@article{Khan2019,
  author = {Khan, Nazir and Prishchenko, Danil and Skourski, Yurii and Mazurenko, Vladimir G. and Tsirlin, Alexander A.},
  title = {{Cubic symmetry and magnetic frustration on the fcc spin lattice in K$_2$IrCl$_6$}},
  journal = {Physical Review B},
  year = {2019},
  volume = {99},
  pages = {144425},
  doi = {10.1103/PhysRevB.99.144425},
  number = {14},
}

@article{ReigiPlessis2020,
  author = {Reig-i-Plessis, D. and Johnson, T. A. and Lu, K. and Chen, Q. and Ruff, J. P. C. and Upton, M. H. and Williams, T. J. and Calder, S. and Zhou, H. D. and Clancy, J. P. and Aczel, A. A. and MacDougall, G. J.},
  title = {{Structural, electronic, and magnetic properties of nearly ideal $J_\mathrm{eff}=1/2$ iridium halides}},
  journal = {Physical Review Materials},
  year = {2020},
  volume = {4},
  pages = {124407},
  doi = {10.1103/PhysRevMaterials.4.124407},
  number = {12},
}

@article{Wang2025,
  author = {Wang, Qiaochu and de la Torre, Alberto and Rodriguez-Rivera, Jose A. and Podlesnyak, Andrey A. and Tian, Wei and Aczel, Adam A. and Matsuda, Masaaki and Ryan, Philip J. and Kim, Jong-Woo and Rau, Jeffrey G. and Plumb, Kemp W.},
  title = {{Pulling Order Back from the Brink of Disorder: Observation of a Nodal-Line Spin Liquid and Fluctuation Stabilized Order in K$_2$IrCl$_6$}},
  journal = {Physical Review X},
  year = {2025},
  volume = {15},
  pages = {021021},
  doi = {10.1103/PhysRevX.15.021021},
  number = {2},
}

@article{Khan2021,
  author = {Khan, Nazir and Prishchenko, Danil and Upton, Mary H. and Mazurenko, Vladimir G. and Tsirlin, Alexander A.},
  title = {{Towards cubic symmetry for Ir$^{4+}$: Structure and magnetism of the antifluorite K$_2$IrBr$_6$}},
  journal = {Physical Review B},
  year = {2021},
  volume = {103},
  pages = {125158},
  doi = {10.1103/PhysRevB.103.125158},
  number = {12},
}

@article{Potzel1977,
  author = {Potzel, W. and Wagner, F.E. and Gierisch, W. and Gebauer, E. and Kalvius, G.M.},
  title = {{M{\"o}ssbauer investigation of the magnetic behaviour of hexahalides of Ir(IV) at very low temperatures}},
  journal = {Physica B+C},
  year = {1977},
  volume = {86-88},
  pages = {899--900},
  doi = {10.1016/0378-4363(77)90733-1},
}

@article{BarandiaranSeijo1988,
  author = {Barandiar{\'a}n, Zoila and Seijo, Luis},
  title = {{The \textit{ab initio} model potential representation of the crystalline environment. Theoretical study of the local distortion on NaCl:Cu$^+$}},
  journal = {The Journal of Chemical Physics},
  year = {1988},
  volume = {89},
  pages = {5739--5746},
  doi = {10.1063/1.455549},
  number = {9},
}

@article{Huzinaga1987,
  author = {Huzinaga, Sigeru and Seijo, Luis and Barandiar{\'a}n, Zoila and Klobukowski, Mariusz},
  title = {{The \textit{ab initio} model potential method. Main group elements}},
  journal = {The Journal of Chemical Physics},
  year = {1987},
  volume = {86},
  pages = {2132--2145},
  doi = {10.1063/1.452111},
  number = {4},
}

@article{Roos1980,
  author = {Roos, Bj{\"o}rn O. and Taylor, Peter R. and Siegbahn, Per E. M.},
  title = {{A complete active space SCF method (CASSCF) using a density matrix formulated super-CI approach}},
  journal = {Chemical Physics},
  year = {1980},
  volume = {48},
  pages = {157--173},
  doi = {10.1016/0301-0104(80)80045-0},
  number = {2},
}

@article{Andersson1990,
  author = {Andersson, Kerstin. and Malmqvist, Per Aake. and Roos, Bjoern O. and Sadlej, Andrzej J. and Wolinski, Krzysztof.},
  title = {{Second-order perturbation theory with a CASSCF reference function}},
  journal = {The Journal of Physical Chemistry},
  year = {1990},
  volume = {94},
  pages = {5483--5488},
  doi = {10.1021/j100377a012},
  number = {14},
}

@article{Andersson1992,
  author = {Andersson, Kerstin and Malmqvist, Per-{\AA}ke and Roos, Bj{\"o}rn O.},
  title = {{Second-order perturbation theory with a complete active space self-consistent field reference function}},
  journal = {The Journal of Chemical Physics},
  year = {1992},
  volume = {96},
  pages = {1218--1226},
  doi = {10.1063/1.462209},
  number = {2},
}

@article{Finley1998,
  author = {Finley, James and Malmqvist, Per-{\AA}ke and Roos, Bj{\"o}rn O. and Serrano-Andr{\'e}s, Luis},
  title = {{The multi-state CASPT$_{2}$ method}},
  journal = {Chemical Physics Letters},
  year = {1998},
  volume = {288},
  pages = {299--306},
  doi = {10.1016/S0009-2614(98)00252-8},
  number = {2-4},
}

@article{Shiozaki2011,
  author = {Shiozaki, Toru and Gy{\H o}rffy, Werner and Celani, Paolo and Werner, Hans-Joachim},
  title = {{Communication: Extended multi-state complete active space second-order perturbation theory: Energy and nuclear gradients}},
  journal = {The Journal of Chemical Physics},
  year = {2011},
  volume = {135},
  pages = {081106},
  doi = {10.1063/1.3633329},
  number = {8},
}

@article{Granovsky2011,
  author = {Granovsky, Alexander A.},
  title = {{Extended multi-configuration quasi-degenerate perturbation theory: The new approach to multi-state multi-reference perturbation theory}},
  journal = {The Journal of Chemical Physics},
  year = {2011},
  volume = {134},
  pages = {214113},
  doi = {10.1063/1.3596699},
  number = {21},
}

@article{Ghigo2004,
  author = {Ghigo, Giovanni and Roos, Bj{\"o}rn O. and Malmqvist, Per-{\AA}ke},
  title = {{A modified definition of the zeroth-order Hamiltonian in multiconfigurational perturbation theory (CASPT$_{2}$)}},
  journal = {Chemical Physics Letters},
  year = {2004},
  volume = {396},
  pages = {142--149},
  doi = {10.1016/j.cplett.2004.08.032},
  number = {1-3},
}

@article{Zobel2017,
  author = {Zobel, J. Patrick and Nogueira, Juan J. and Gonz{\'a}lez, Leticia},
  title = {{The IPEA dilemma in CASPT$_{2}$}},
  journal = {Chemical Science},
  year = {2017},
  volume = {8},
  pages = {1482--1499},
  doi = {10.1039/C6SC03759C},
  number = {2},
}

@article{Battaglia2020,
  author = {Battaglia, Stefano and Lindh, Roland},
  title = {{Extended Dynamically Weighted CASPT$_{2}$: The Best of Two Worlds}},
  journal = {Journal of Chemical Theory and Computation},
  year = {2020},
  volume = {16},
  pages = {1555--1567},
  doi = {10.1021/acs.jctc.9b01129},
  number = {3},
}

@article{Nishimoto2023,
  author = {Nishimoto, Yoshio},
  title = {{Analytic first-order derivatives of CASPT$_{2}$ with IPEA shift}},
  journal = {The Journal of Chemical Physics},
  year = {2023},
  volume = {158},
  pages = {174112},
  doi = {10.1063/5.0147611},
  number = {17},
}

@article{Angeli2001,
  author = {Angeli, C. and Cimiraglia, R. and Evangelisti, S. and Leininger, T. and Malrieu, J.-P.},
  title = {{Introduction of $n$-electron valence states for multireference perturbation theory}},
  journal = {The Journal of Chemical Physics},
  year = {2001},
  volume = {114},
  pages = {10252--10264},
  doi = {10.1063/1.1361246},
  number = {23},
}

@article{Angeli2002,
  author = {Angeli, Celestino and Cimiraglia, Renzo and Malrieu, Jean-Paul},
  title = {{$n$-electron valence state perturbation theory: A spinless formulation and an efficient implementation of the strongly contracted and of the partially contracted variants}},
  journal = {The Journal of Chemical Physics},
  year = {2002},
  volume = {117},
  pages = {9138--9153},
  doi = {10.1063/1.1515317},
  number = {20},
}

@article{Angeli2004,
  author = {Angeli, Celestino and Borini, Stefano and Cestari, Mirko and Cimiraglia, Renzo},
  title = {{A quasidegenerate formulation of the second order $n$-electron valence state perturbation theory approach}},
  journal = {The Journal of Chemical Physics},
  year = {2004},
  volume = {121},
  pages = {4043--4049},
  doi = {10.1063/1.1778711},
  number = {9},
}

@article{Lang2019,
  author = {Lang, Lucas and Neese, Frank},
  title = {{Spin-dependent properties in the framework of the dynamic correlation dressed complete active space method}},
  journal = {The Journal of Chemical Physics},
  year = {2019},
  volume = {150},
  pages = {104104},
  doi = {10.1063/1.5085203},
  number = {10},
}

@article{Malmqvist2002,
  author = {Malmqvist, Per {\AA}ke and Roos, Bj{\"o}rn O. and Schimmelpfennig, Bernd},
  title = {{The restricted active space (RAS) state interaction approach with spin--orbit coupling}},
  journal = {Chemical Physics Letters},
  year = {2002},
  volume = {357},
  pages = {230--240},
  doi = {10.1016/S0009-2614(02)00498-0},
  number = {3-4},
}

@article{Hess1996,
  author = {He{\ss}, Bernd A. and Marian, Christel M. and Wahlgren, Ulf and Gropen, Odd},
  title = {{A mean-field spin-orbit method applicable to correlated wavefunctions}},
  journal = {Chemical Physics Letters},
  year = {1996},
  volume = {251},
  pages = {365--371},
  doi = {10.1016/0009-2614(96)00119-4},
  number = {5-6},
}

@article{Hess1986,
  author = {Hess, Bernd A.},
  title = {{Relativistic electronic-structure calculations employing a two-component no-pair formalism with external-field projection operators}},
  journal = {Physical Review A},
  year = {1986},
  volume = {33},
  pages = {3742--3748},
  doi = {10.1103/PhysRevA.33.3742},
  number = {6},
}

@article{Roos2004,
  author = {Roos, Bj{\"o}rn O. and Lindh, Roland and Malmqvist, Per-{\AA}ke and Veryazov, Valera and Widmark, Per-Olof},
  title = {{Main Group Atoms and Dimers Studied with a New Relativistic ANO Basis Set}},
  journal = {The Journal of Physical Chemistry A},
  year = {2004},
  volume = {108},
  pages = {2851--2858},
  doi = {10.1021/jp031064+},
  number = {15},
}

@article{Roos2005,
  author = {Roos, Bj{\"o}rn O. and Lindh, Roland and Malmqvist, Per-{\AA}ke and Veryazov, Valera and Widmark, Per-Olof},
  title = {{New Relativistic ANO Basis Sets for Transition Metal Atoms}},
  journal = {The Journal of Physical Chemistry A},
  year = {2005},
  volume = {109},
  pages = {6575--6579},
  doi = {10.1021/jp0581126},
  number = {29},
}

@article{FdezGalvan2019,
  author = {Fdez. Galv{\'a}n, Ignacio and Vacher, Morgane and Alavi, Ali and Angeli, Celestino and Aquilante, Francesco and Autschbach, Jochen and Bao, Jie J. and Bokarev, Sergey I. and Bogdanov, Nikolay A. and Carlson, Rebecca K. and Chibotaru, Liviu F. and Creutzberg, Joel and Dattani, Nike and Delcey, Micka{\"e}l G. and Dong, Sijia S. and Dreuw, Andreas and Freitag, Leon and Frutos, Luis Manuel and Gagliardi, Laura and Gendron, Fr{\'e}d{\'e}ric and Giussani, Angelo and Gonz{\'a}lez, Leticia and Grell, Gilbert and Guo, Meiyuan and Hoyer, Chad E. and Johansson, Marcus and Keller, Sebastian and Knecht, Stefan and Kova{\v c}evi{\'c}, Goran and K{\"a}llman, Erik and Li Manni, Giovanni and Lundberg, Marcus and Ma, Yingjin and Mai, Sebastian and Malhado, Jo{\~a}o Pedro and Malmqvist, Per {\AA}ke and Marquetand, Philipp and Mewes, Stefanie A. and Norell, Jesper and Olivucci, Massimo and Oppel, Markus and Phung, Quan Manh and Pierloot, Kristine and Plasser, Felix and Reiher, Markus and Sand, Andrew M. and Schapiro, Igor and Sharma, Prachi and Stein, Christopher J. and S{\o}rensen, Lasse Kragh and Truhlar, Donald G. and Ugandi, Mihkel and Ungur, Liviu and Valentini, Alessio and Vancoillie, Steven and Veryazov, Valera and Weser, Oskar and Weso{\l}owski, Tomasz A. and Widmark, Per-Olof and Wouters, Sebastian and Zech, Alexander and Zobel, J. Patrick and Lindh, Roland},
  title = {{OpenMolcas: From Source Code to Insight}},
  journal = {Journal of Chemical Theory and Computation},
  year = {2019},
  volume = {15},
  pages = {5925--5964},
  doi = {10.1021/acs.jctc.9b00532},
  number = {11},
}

@article{LiManni2023,
  author = {Li Manni, Giovanni and Fdez. Galv{\'a}n, Ignacio and Alavi, Ali and Aleotti, Flavia and Aquilante, Francesco and Autschbach, Jochen and Avagliano, Davide and Baiardi, Alberto and Bao, Jie J. and Battaglia, Stefano and Birnoschi, Letitia and Blanco-Gonz{\'a}lez, Alejandro and Bokarev, Sergey I. and Broer, Ria and Cacciari, Roberto and Calio, Paul B. and Carlson, Rebecca K. and Carvalho Couto, Rafael and Cerd{\'a}n, Luis and Chibotaru, Liviu F. and Chilton, Nicholas F. and Church, Jonathan Richard and Conti, Irene and Coriani, Sonia and Cu{\'e}llar-Zuquin, Juliana and Daoud, Razan E. and Dattani, Nike and Decleva, Piero and de Graaf, Coen and Delcey, Micka{\"e}l G. and De Vico, Luca and Dobrautz, Werner and Dong, Sijia S. and Feng, Rulin and Ferr{\'e}, Nicolas and Filatov(Gulak), Michael and Gagliardi, Laura and Garavelli, Marco and Gonz{\'a}lez, Leticia and Guan, Yafu and Guo, Meiyuan and Hennefarth, Matthew R. and Hermes, Matthew R. and Hoyer, Chad E. and Huix-Rotllant, Miquel and Jaiswal, Vishal Kumar and Kaiser, Andy and Kaliakin, Danil S. and Khamesian, Marjan and King, Daniel S. and Kochetov, Vladislav and Kro{\'s}nicki, Marek and Kumaar, Arpit Arun and Larsson, Ernst D. and Lehtola, Susi and Lepetit, Marie-Bernadette and Lischka, Hans and L{\'o}pez R{\'i}os, Pablo and Lundberg, Marcus and Ma, Dongxia and Mai, Sebastian and Marquetand, Philipp and Merritt, Isabella C. D. and Montorsi, Francesco and M{\"o}rchen, Maximilian and Nenov, Artur and Nguyen, Vu Ha Anh and Nishimoto, Yoshio and Oakley, Meagan S. and Olivucci, Massimo and Oppel, Markus and Padula, Daniele and Pandharkar, Riddhish and Phung, Quan Manh and Plasser, Felix and Raggi, Gerardo and Rebolini, Elisa and Reiher, Markus and Rivalta, Ivan and Roca-Sanju{\'a}n, Daniel and Romig, Thies and Safari, Arta Anushirwan and S{\'a}nchez-Mansilla, Aitor and Sand, Andrew M. and Schapiro, Igor and Scott, Thais R. and Segarra-Mart{\'i}, Javier and Segatta, Francesco and Sergentu, Dumitru-Claudiu and Sharma, Prachi and Shepard, Ron and Shu, Yinan and Staab, Jakob K. and Straatsma, Tjerk P. and S{\o}rensen, Lasse Kragh and Tenorio, Bruno Nunes Cabral and Truhlar, Donald G. and Ungur, Liviu and Vacher, Morgane and Veryazov, Valera and Vo{\ss}, Torben Arne and Weser, Oskar and Wu, Dihua and Yang, Xuchun and Yarkony, David and Zhou, Chen and Zobel, J. Patrick and Lindh, Roland},
  title = {{The OpenMolcas Web : A Community-Driven Approach to Advancing Computational Chemistry}},
  journal = {Journal of Chemical Theory and Computation},
  year = {2023},
  volume = {19},
  pages = {6933--6991},
  doi = {10.1021/acs.jctc.3c00182},
  number = {20},
}

@article{Neese2025,
  author = {Neese, Frank},
  title = {{Software Update: The ORCA Program System---Version 6.0}},
  journal = {WIREs Computational Molecular Science},
  year = {2025},
  volume = {15},
  pages = {e70019},
  doi = {10.1002/wcms.70019},
  number = {2},
}

@article{Olsen1988,
  author = {Olsen, Jeppe and Roos, Bj{\"o}rn O. and J{\o}rgensen, Poul and Jensen, Hans J{\o}rgen Aa.},
  title = {{Determinant based configuration interaction algorithms for complete and restricted configuration interaction spaces}},
  journal = {The Journal of Chemical Physics},
  year = {1988},
  volume = {89},
  pages = {2185--2192},
  doi = {10.1063/1.455063},
  number = {4},
}

@article{Kubo1957,
  author = {Kubo, Ryogo},
  title = {{Statistical-Mechanical Theory of Irreversible Processes. I. General Theory and Simple Applications to Magnetic and Conduction Problems}},
  journal = {Journal of the Physical Society of Japan},
  year = {1957},
  volume = {12},
  pages = {570--586},
  doi = {10.1143/JPSJ.12.570},
  number = {6},
}

@article{Shiozaki2018,
  author = {Shiozaki, Toru},
  title = {{BAGEL: Brilliantly Advanced General Electronic-structure Library}},
  journal = {WIREs Computational Molecular Science},
  year = {2018},
  volume = {8},
  pages = {e1331},
  doi = {10.1002/wcms.1331},
  number = {1},
}

@article{Vlaisavljevich2016,
  author = {Vlaisavljevich, Bess and Shiozaki, Toru},
  title = {{Nuclear Energy Gradients for Internally Contracted Complete Active Space Second-Order Perturbation Theory: Multistate Extensions}},
  journal = {Journal of Chemical Theory and Computation},
  year = {2016},
  volume = {12},
  pages = {3781--3787},
  doi = {10.1021/acs.jctc.6b00572},
  number = {8},
}

@article{MacLeod2015,
  author = {MacLeod, Matthew K. and Shiozaki, Toru},
  title = {{Communication: Automatic code generation enables nuclear gradient computations for fully internally contracted multireference theory}},
  journal = {The Journal of Chemical Physics},
  year = {2015},
  volume = {142},
  pages = {051103},
  doi = {10.1063/1.4907717},
  number = {5},
}

@article{Park2020,
  author = {Park, Jae Woo and Al-Saadon, Rachael and MacLeod, Matthew K. and Shiozaki, Toru and Vlaisavljevich, Bess},
  title = {{Multireference Electron Correlation Methods: Journeys along Potential Energy Surfaces}},
  journal = {Chemical Reviews},
  year = {2020},
  volume = {120},
  pages = {5878--5909},
  doi = {10.1021/acs.chemrev.9b00496},
  number = {13},
}

@misc{CompanionMechanism,
  author = {Y{\i}ld{\i}r{\i}m, Y{\"u}cel},
  title = {Crystal-controlled fine structure of spin-orbit excitons in {LiIrF$_6$} and {CsIrF$_6$}},
  note = {unpublished companion manuscript},
  year = {2026}
}

@article{Safari2024,
  author = {Safari, Arta A. and Anderson, Robert J. and Li Manni,
    Giovanni},
  title = {Toward a {Stochastic} {Complete} {Active} {Space}
    {Second-Order} {Perturbation} {Theory}},
  journal = {The Journal of Physical Chemistry A},
  volume = {128},
  number = {1},
  pages = {281-291},
  year = {2024},
  month = {jan},
  url = {https://doi.org/10.1021/acs.jpca.3c05109}
}

@article{Park2019,
  author = {Park, Jae Woo},
  title = {Single-State Single-Reference and Multistate Multireference Zeroth-Order Hamiltonians in {MS-CASPT2} and Conical Intersections},
  journal = {Journal of Chemical Theory and Computation},
  year = {2019}, volume = {15}, number = {7}, pages = {3960--3973},
  doi = {10.1021/acs.jctc.9b00067}
}

@article{BattagliaLindh2021,
  author = {Battaglia, Stefano and Lindh, Roland},
  title = {On the role of symmetry in {XDW-CASPT2}},
  journal = {The Journal of Chemical Physics},
  year = {2021}, volume = {154}, number = {3}, pages = {034102},
  doi = {10.1063/5.0030944}
}

\end{document}

\maketitle

\section{S1. Cluster definitions, active spaces and native
spectra}\label{s1.-cluster-definitions-active-spaces-and-native-spectra}

The eight compounds span nominal Ir(V)/Os(IV) \(d^4\) and Ir(IV) \(d^5\)
configurations. Formal valences specify electron bookkeeping rather than
fitted atomic charges. Each metal-shell/ligand-active pair retains its
own geometry, basis, scalar-relativistic treatment and embedding. This
controls a comparison within one material; it does not establish equal
boundary accuracy across the family. The Ba\(_2\)YIrO\(_6\) calculations
reported here use Y(III) AIMPs from the Cs\(_2\)NaYF\(_6\) library entry
and Ba(II) AIMPs from BaF\(_2\). Transfer from those host environments
remains an embedding approximation.

Baseline calculations use ANO-RCC-VDZP on the central metal and six
ligands, except Ir/O in Ba\(_2\)YIrO\(_6\), which use VTZP; scalar
relativity is DKH2\textsuperscript{1--3}. The five-orbital metal shell
contains four or five electrons. Ligand-active references add five
initially doubly occupied combinations, producing CAS(14,10) or
CAS(15,10). The complete metal-shell root counts are 50/45/5
singlets/triplets/quintets for \(d^4\) and 75/24/1
doublets/quartets/sextets for \(d^5\). The retained ligand-active counts
are:

{\def\LTcaptype{none} % do not increment counter
\begin{longtable}[]{@{}
  >{\raggedright\arraybackslash}p{(\linewidth - 6\tabcolsep) * \real{0.2500}}
  >{\raggedright\arraybackslash}p{(\linewidth - 6\tabcolsep) * \real{0.2500}}
  >{\raggedright\arraybackslash}p{(\linewidth - 6\tabcolsep) * \real{0.2500}}
  >{\raggedright\arraybackslash}p{(\linewidth - 6\tabcolsep) * \real{0.2500}}@{}}
\toprule\noalign{}
\begin{minipage}[b]{\linewidth}\raggedright
Compound
\end{minipage} & \begin{minipage}[b]{\linewidth}\raggedright
Roots in increasing spin multiplicity
\end{minipage} & \begin{minipage}[b]{\linewidth}\raggedright
MS metal / ligand centroid
\end{minipage} & \begin{minipage}[b]{\linewidth}\raggedright
XMS metal / ligand centroid
\end{minipage} \\
\midrule\noalign{}
\endhead
\bottomrule\noalign{}
\endlastfoot
LiIrF\(_6\) & 33 / 32 / 15 & 527.522 / 460.591 & 533.900 / 458.073 \\
Ba\(_2\)YIrO\(_6\) & 31 / 33 / 15 & 443.999 / 387.477 & 463.895 /
397.959 \\
K\(_2\)OsCl\(_6\) & 34 / 32 / 17 & 361.956 / 350.818 & 372.703 /
341.469 \\
K\(_2\)OsBr\(_6\) & 34 / 32 / 17 & 369.361 / 347.644 & 376.280 /
341.443 \\
K\(_2\)IrF\(_6\) & 65 / 32 / 3 & 811.675 / 779.111 & 798.356 /
778.970 \\
Cs\(_2\)IrF\(_6\) & 65 / 34 / 3 & 820.464 / 799.738 & 807.820 /
789.415 \\
K\(_2\)IrCl\(_6\) & 54 / 30 / 1 & 676.972 / 659.113 & 666.582 /
651.693 \\
K\(_2\)IrBr\(_6\) & 44 / 29 / 1 & 684.561 / 645.380 & 675.725 /
643.294 \\
\end{longtable}
}

Table S1. Raw, unprojected SOC first-excitation centroids in meV for the
listed root windows. Each excitation is measured from the ground state
of its own operator. The first four rows use singlet/triplet/quintet
roots; the others use doublet/quartet/sextet roots. Each difference is
evaluated before rounding. MS and XMS are different protocols, not error
bars. Cs\(_2\)IrF\(_6\) is Ir(IV), distinct from the auxiliary Ir(V)
CsIrF\(_6\) model.

The accompanying
\texttt{01\_calculation\_records/baseline\_inputs/index.json} identifies
all 96 native MS/XMS input decks and their source-output records. It
provides the full Cartesian coordinates, explicit-ion and
boundary-potential labels, resolved basis labels, root counts,
inactive/frozen-orbital records and file hashes. All referenced orbital
and point-field input files are included; Table S1b summarizes the
native electron counts. The installed OpenMolcas basis/potential library
remains an external dependency; the original Y/Ba host-potential labels
are preserved in the native input index. The corresponding native
outputs and RASSI HDF5 files are included under
\texttt{01\_calculation\_records/baseline\_native}, together with a
spectrum-reconstruction script.

\Needspace{19\baselineskip}

{\def\LTcaptype{none} % do not increment counter
\begin{longtable}[]{@{}
  >{\raggedright\arraybackslash}p{(\linewidth - 8\tabcolsep) * \real{0.2000}}
  >{\raggedleft\arraybackslash}p{(\linewidth - 8\tabcolsep) * \real{0.2000}}
  >{\raggedright\arraybackslash}p{(\linewidth - 8\tabcolsep) * \real{0.2000}}
  >{\raggedleft\arraybackslash}p{(\linewidth - 8\tabcolsep) * \real{0.2000}}
  >{\raggedleft\arraybackslash}p{(\linewidth - 8\tabcolsep) * \real{0.2000}}@{}}
\toprule\noalign{}
\begin{minipage}[b]{\linewidth}\raggedright
Compound
\end{minipage} & \begin{minipage}[b]{\linewidth}\raggedleft
Electrons
\end{minipage} & \begin{minipage}[b]{\linewidth}\raggedright
Inactive metal / ligand
\end{minipage} & \begin{minipage}[b]{\linewidth}\raggedleft
PT2 frozen
\end{minipage} & \begin{minipage}[b]{\linewidth}\raggedleft
Boundary centres
\end{minipage} \\
\midrule\noalign{}
\endhead
\bottomrule\noalign{}
\endlastfoot
Ba\(_2\)YIrO\(_6\) & 132 & 64 / 59 & 36 & 14 \\
Cs\(_2\)IrF\(_6\) & 133 & 64 / 59 & 36 & 12 \\
K\(_2\)IrBr\(_6\) & 289 & 142 / 137 & 114 & 8 \\
K\(_2\)IrCl\(_6\) & 181 & 88 / 83 & 60 & 8 \\
K\(_2\)IrF\(_6\) & 133 & 64 / 59 & 36 & 12 \\
K\(_2\)OsBr\(_6\) & 288 & 142 / 137 & 114 & 8 \\
K\(_2\)OsCl\(_6\) & 180 & 88 / 83 & 60 & 8 \\
LiIrF\(_6\) & 144 & 70 / 65 & 42 & 86 \\
\end{longtable}
}

Table S1b. Counts read from the native inputs/outputs. Inactive counts
are before CASPT2 freezing; the frozen count is identical for both
spaces. Boundary centres exclude the central metal and six ligands; Li
includes six explicit all-electron Li centres plus the frozen-ion
boundary. Embedded input charge keywords include the potential
convention and should not be interpreted as isolated-cluster charges.
Coordinates are the actual fixed calculation inputs, not new
force-converged relaxations.

A separate paired VTZP control compares the physical Y/Ba AIMP boundary
with the historical In/Ba ECP surrogate. Geometry, point charges, basis
and the 45-triplet orbital-optimization ensemble are held to the same
protocol; each boundary has its own optimized orbitals, and CAS/2h0p
within each boundary share their orbital-selection hash.

{\def\LTcaptype{none} % do not increment counter
\begin{longtable}[]{@{}
  >{\raggedright\arraybackslash}p{(\linewidth - 6\tabcolsep) * \real{0.2500}}
  >{\raggedleft\arraybackslash}p{(\linewidth - 6\tabcolsep) * \real{0.2500}}
  >{\raggedleft\arraybackslash}p{(\linewidth - 6\tabcolsep) * \real{0.2500}}
  >{\raggedleft\arraybackslash}p{(\linewidth - 6\tabcolsep) * \real{0.2500}}@{}}
\toprule\noalign{}
\begin{minipage}[b]{\linewidth}\raggedright
Boundary
\end{minipage} & \begin{minipage}[b]{\linewidth}\raggedleft
CAS centroid (meV)
\end{minipage} & \begin{minipage}[b]{\linewidth}\raggedleft
2h0p centroid (meV)
\end{minipage} & \begin{minipage}[b]{\linewidth}\raggedleft
2h0p minus CAS (meV)
\end{minipage} \\
\midrule\noalign{}
\endhead
\bottomrule\noalign{}
\endlastfoot
Y/Ba AIMP & 480.124 & 464.419 & -15.7044 \\
In/Ba ECP & 490.111 & 475.674 & -14.4362 \\
\end{longtable}
}

Table S1c. Paired oxide-boundary sensitivity control. Differences use
unrounded source energies: replacing In/Ba ECP by Y/Ba AIMP lowers CAS
and 2h0p centroids by 9.9869 and 11.2551 meV. Each final calculation
retains 5/45/50 quintet/triplet/singlet roots. The ligand-active windows
omit spin sectors beyond quintets; the Y potential is host transferred.
The comparison bounds this protocol and does not supply a correction to
the separate ORCA model or the baseline magnetic-response table.

For Li, the central bare anion has 132 electrons. The embedded
OpenMolcas cluster includes six explicit closed-shell Li\(^+\) cores and
has 144 electrons. Its metal-shell reference has 70 inactive orbitals;
the ligand-active reference has 65. Frozen-ion potentials do not add
variational electrons. Their effective nuclear charges determine the
program's input charge, which need not equal the physical charge of the
central anion.

Occupied ligand combinations are defined in the AO metric. If \(C_B\)
contains eighteen ligand-\(p\) parent orbitals, \(S\) is the overlap and
\(P_d\) selects metal-\(d\) rows in the orthonormal AO frame, the SVD

\begin{equation}P_dS^{1/2}C_B=W\Sigma V^\dagger,\qquad C_L=C_BV_{[:,1:5]}\end{equation}

defines the selected five-dimensional subspace. Its retained weight is
\(\sum_{i=1}^{5}\sigma_i^2/\sum_i\sigma_i^2\), 99.981--99.998\% in the
tested fluoride controls. This fraction refers to the specified
projector, not all ligand correlation. The singular-value boundary and
principal angles against alternative projections are
checked\textsuperscript{4}. The ligand population of state \(n\) is
\(N_L(n)=\operatorname{Tr}(C_L^\dagger S D_n S C_L)\), with hole count
\(10-N_L\). Sorted natural occupations alone do not specify this
subspace.

\begin{figure}
\centering
\pandocbounded{\includegraphics[keepaspectratio,alt={All 32 native first-manifold spectra. Ticks represent stored SOC eigenvalues, with no displacement of overlapping levels; symbols mark the arithmetic centroids. Material panels use their own scales. The width of a noncubic manifold is not automatically a numerical symmetry defect. Experimental comparisons in the main text retain their technique and estimator.}]{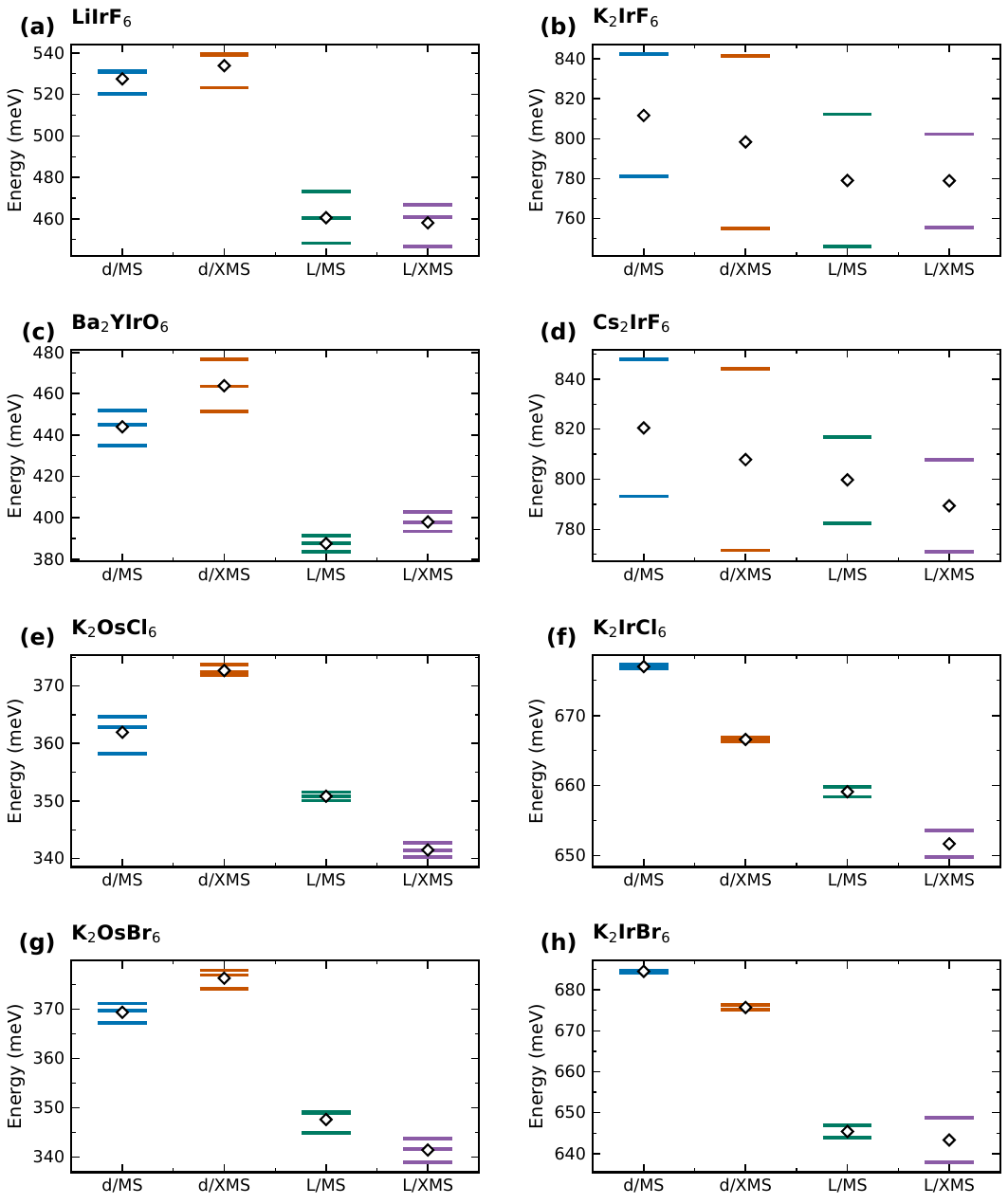}}
\caption{All 32 native first-manifold spectra. Ticks represent stored
SOC eigenvalues, with no displacement of overlapping levels; symbols
mark the arithmetic centroids. Material panels use their own scales. The
width of a noncubic manifold is not automatically a numerical symmetry
defect. Experimental comparisons in the main text retain their technique
and estimator.}
\end{figure}

\section{S2. Effective Hamiltonian
reconstruction}\label{s2.-effective-hamiltonian-reconstruction}

The reconstruction uses native CASCI energies, the MS/XMS effective
Hamiltonians, reference-state transformations and SOC
matrices\textsuperscript{5,6}. All terms must be expressed in the same
ordered reference gauge. If columns of \(V\) express the rotated XMS
references in the original basis,
\(H^{\rm CASCI}=VH^{\rm XMS}V^\dagger\). For a rounded transformation
\(V=A\Sigma B^\dagger\), its closest orthogonal polar factor is
\(AB^\dagger\). We apply this to the small relative Hamiltonian after
separating a common scalar energy; that energy is restored when
comparing absolute values.

In the basis \(|a,S,M\rangle\),

\begin{equation}\langle a,S,M|H_{\rm sf}^{\rm PT2}|b,S',M'\rangle=\delta_{SS'}\delta_{MM'}(H_S^{\rm PT2})_{ab},\end{equation}

\begin{equation}H_{\rm eff}=\bigoplus_S(H_S^{\rm PT2}\otimes I_{2S+1})+H_{\rm SO}.\end{equation}

The spin-free block includes reference energies and their perturbative
dressing. It is not a correction to be added twice. The index \(a\)
labels a reference root, \(S\) total spin and \(M\) its projection. The
direct sum assembles sectors; the tensor product repeats each matrix
element across the magnetic sublevels. A complete \(d^4\) metal-shell
space has dimension \(50+45\times3+5\times5=210\). SOC can mix roots and
sectors where allowed.

Hermiticity, absolute offsets and excitation differences are tested
separately. Missing a common energy offset can leave every excitation
unchanged. Across 32 baseline spectra the maximum reconstruction
discrepancy is 0.000247921 meV. A broader independent diagonalization of
56 saved SOC matrices reproduces relative levels within
\(6.8\times10^{-14}\) Hartree and gives Hermiticity errors below
\(2.1\times10^{-17}\) Hartree. These are numerical consistency bounds
for saved matrices; they do not bound external correlation or states
absent from the window.

An independent fermionic operator implementation also checks variational
SOC and magnetic conventions. Its Hamiltonian action agrees near
\(10^{-10}\) Hartree, and 488 available native SOC levels agree within
0.003287 meV. First-manifold magnetic strength agrees within
\(1.6\times10^{-7}\,\mu_B^2\). This independent kernel is not another
implementation of MS/XMS-CASPT2. Matrix and numerical-result checks are
supplied separately in the reproducibility files so their scopes remain
explicit.

The per-case 32-spectrum replay is recorded in
\texttt{baseline32\_matrix\_replay.json} in the source-records folder;
\texttt{02\_validation\_scripts/replay\_baseline32.py} repeats it from
the included native archive. This is a matrix reconstruction check,
distinct from rereading stored eigenvalues to recompute centroids. All
archived baseline banners identify OpenMolcas 24.02.

\section{S3. Embedded Li symmetry and perturbative
controls}\label{s3.-embedded-li-symmetry-and-perturbative-controls}

For an AO rotation \(D\) and an \(S\)-orthonormal orbital frame \(C\),
the induced subspace operation is \(U=C^\dagger SDC\). The subspace
leakage is measured by the largest eigenvalue of
\((DC-CU)^\dagger S(DC-CU)\). This is a squared measure. In the original
Li reference it is \(3.60\times10^{-8}\) for the active projector.
Symmetry-related fitted charges differ by at most 0.00361834 electron.
Charge and projector symmetrization are therefore separate
interventions.

Charge averaging over geometrical threefold orbits and projector
averaging/purification retain the electron count and orbital ranks. The
active-projector squared leakage falls to \(1.83\times10^{-19}\), and MO
orthonormality remains within \(6.55\times10^{-14}\). The largest
projector displacement is approximately \(1.1\times10^{-4}\). No
excitation energy is used to construct the modified reference.

{\def\LTcaptype{none} % do not increment counter
\begin{longtable}[]{@{}lrr@{}}
\toprule\noalign{}
Reference & Final-CI pair (meV) & Density covariance residual \\
\midrule\noalign{}
\endhead
\bottomrule\noalign{}
\endlastfoot
Charges symmetrized & 0.144542 & \(8.36\times10^{-6}\) \\
Orbital subspaces symmetrized & 0.000495 & \(2.90\times10^{-7}\) \\
Both changes & 0.000293 & \(2.95\times10^{-8}\) \\
\end{longtable}
}

Table S2. Reference-stage tests. The density residual is a
matrix-element difference, not an energy uncertainty. Final-CI energies
are used because an earlier saved root array differs by up to 0.033 meV
per state.

{\def\LTcaptype{none} % do not increment counter
\begin{longtable}[]{@{}lrr@{}}
\toprule\noalign{}
Li protocol & SOC centroid (meV) & Closest excited pair (meV) \\
\midrule\noalign{}
\endhead
\bottomrule\noalign{}
\endlastfoot
Original, MS, IPEA 0.25 & 460.591 & 11.990 \\
Original, XMS, IPEA 0.25 & 458.073 & 5.735 \\
Original, XMS, canonical orthogonalization & 458.012 & 5.728 \\
Original, XMS, IPEA 0 & 440.469 & 0.367 \\
Both reference changes, XMS, IPEA 0.25 & 458.600 & 5.251 \\
Both reference changes, XMS, IPEA 0 & 440.446 & 0.600 \\
\end{longtable}
}

Table S3. Spin--orbit results, distinct from Table S2's spin-free
reference pairs. IPEA is in Hartree; imaginary shift is 0.1 Hartree
throughout. All inspected perturbative blocks finish normally with
equation residuals at most \(10^{-6}\). A smaller pair defect does not
choose an experimentally more accurate centroid.

\section{S4. Group projection, rotation and root
selection}\label{s4.-group-projection-rotation-and-root-selection}

For the even-electron Li controls, the fitted threefold representation
includes spin action when projecting the full SOC Hamiltonian. The
averaging map is

\begin{equation}\mathcal P_U(H)=\frac13\sum_{k=0}^{2}U^kH(U^\dagger)^k.\end{equation}

A valid finite representation requires \(U^\dagger U=I\) and \(U^3=I\).
Its physical interpretation additionally requires reference commutation
and vector covariance,
\(UA_\alpha U^\dagger=\sum_\beta R_{\alpha\beta}A_\beta\). Dipole
operators withheld from the fit test information not used to determine
\(U\). Threefold symmetry has one-dimensional complex irreducible
representations. The interpretation of an even-electron pair also uses
the applicable real/time-reversal structure, which pairs conjugate
threefold sectors; \(C_3\) alone is not a universal complex-space
doublet-protection argument. This construction is not a full
double-group representation for the odd-electron family.

The original-reference rotation scan and the corrected-reference test
use different fitted representations. In the original MS reference, the
angular-momentum and withheld AMFI residuals are \(9.6110\times10^{-4}\)
and \(1.4178\times10^{-3}\); the four rotations retain these scales. The
corrected embedded reference instead gives angular-momentum/AMFI/dipole
residuals near \(3.2/4.3/2.8\times10^{-6}\). Finite-order correction
reduces the largest element of \(U^3-I\) below \(2.6\times10^{-14}\)
across seventeen cases; centroid changes are below \(3.7\times10^{-7}\)
meV. The corrected-reference projected XMS centroid nevertheless changes
by 16.125 meV between IPEA 0.25 and zero. Projection removes only the
part incompatible with its chosen symmetry action.

For these embedded controls, \(U\) is inferred from reference energy
groups and operator covariance, rather than obtained from a direct
many-electron rotation witness. AMFI is dominated by one-centre Ir
contributions and is partly correlated with the fitted angular momentum;
the separately withheld dipole supplies an additional check. The AO/CI
determinant rotation witness in S5 applies to the independent
three-state bare-anion model and does not validate the embedded
representations.

The four reference orientations give projected MS centroids 456.040,
456.215, 456.060 and 456.093 meV. At zero degrees, the raw centroid is
460.591 meV with a closest-pair defect of 11.990 meV; the projected
centroid is 456.040 meV with an A--E separation of 13.66 meV. These are
different operators in the same 33-singlet/32-triplet/15-quintet SOC
window. These tests rotate reference states, not physical nuclei or
active orbitals. An active-orbital rotation is a distinct operation,
particularly for restricted configuration sectors\textsuperscript{7,8}.
In the embedded root-selection rule, states within \(10^{-5}\) Hartree
form groups. A selected boundary must include a whole energy group as a
practical safeguard. The actual symmetry-closure condition is
\([P,U_g]=0\) for the retained-subspace projector \(P\); an energy-group
boundary alone does not prove this condition. The bare-anion grouping
tolerance is \(2\times10^{-6}\) Hartree, distinct from the embedded
rule.

The larger Li window contains 59 triplets, 58 singlets and 25 quintets.
Its native MS/XMS centroids change by +19.447/+8.525 meV. The projected
changes, +17.518/+10.173 meV, are diagnostic only: the inferred
representation has a 0.599 meV reference commutator against a chosen 0.5
meV criterion, or 1.88\% covariance error against 1\%. These criteria
are stated tests, not confidence intervals.

The following operator comparison uses the angular-momentum-only
least-squares construction in each XMS reference. AMFI and SOC were
withheld from that fit; they are not mutually independent because AMFI
contributes to SOC. These dimensionless residuals are numerical
covariance diagnostics, not energy uncertainties. They are also distinct
from the joint-fit and reference-commutator tests of the enlarged window
discussed above.

{\def\LTcaptype{none} % do not increment counter
\begin{longtable}[]{@{}
  >{\raggedright\arraybackslash}p{(\linewidth - 6\tabcolsep) * \real{0.2500}}
  >{\raggedleft\arraybackslash}p{(\linewidth - 6\tabcolsep) * \real{0.2500}}
  >{\raggedleft\arraybackslash}p{(\linewidth - 6\tabcolsep) * \real{0.2500}}
  >{\raggedleft\arraybackslash}p{(\linewidth - 6\tabcolsep) * \real{0.2500}}@{}}
\toprule\noalign{}
\begin{minipage}[b]{\linewidth}\raggedright
Reference
\end{minipage} & \begin{minipage}[b]{\linewidth}\raggedleft
Fitted \(L\) residual
\end{minipage} & \begin{minipage}[b]{\linewidth}\raggedleft
Withheld AMFI residual
\end{minipage} & \begin{minipage}[b]{\linewidth}\raggedleft
Withheld SOC residual
\end{minipage} \\
\midrule\noalign{}
\endhead
\bottomrule\noalign{}
\endlastfoot
Original & \(9.6110\times10^{-4}\) & \(1.4178\times10^{-3}\) &
\(1.4178\times10^{-3}\) \\
Corrected, IPEA 0.25 & \(3.2365\times10^{-6}\) & \(4.3219\times10^{-6}\)
& \(4.3219\times10^{-6}\) \\
Enlarged root window & \(1.3257\times10^{-3}\) & \(2.0271\times10^{-3}\)
& \(2.0271\times10^{-3}\) \\
\end{longtable}
}

Table S3b. Reference-specific covariance checks from the native operator
records. The original MS values differ from the displayed XMS values
only beyond the shown precision. A small finite-order residual alone
does not certify the physical reference or convergence of its state
window.

\begin{figure}
\centering
\pandocbounded{\includegraphics[keepaspectratio,alt={Embedded Li representation checks. (a) Native and projected MS centroid changes along the original-reference rotation scan, measured relative to the zero-angle result. (b) Angular momentum used in the fit and withheld AMFI/SOC residuals for original, corrected and enlarged XMS references. The corrected-reference residuals must not be assigned to the original rotation scan. Withheld operators provide additional tests, while AMFI and SOC remain correlated.}]{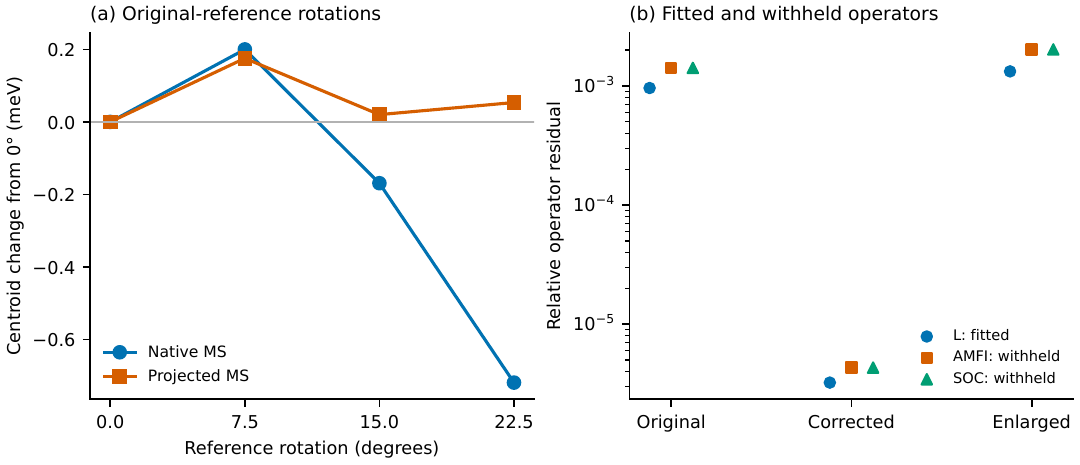}}
\caption{Embedded Li representation checks. (a) Native and projected MS
centroid changes along the original-reference rotation scan, measured
relative to the zero-angle result. (b) Angular momentum used in the fit
and withheld AMFI/SOC residuals for original, corrected and enlarged XMS
references. The corrected-reference residuals must not be assigned to
the original rotation scan. Withheld operators provide additional tests,
while AMFI and SOC remain correlated.}
\end{figure}

\FloatBarrier

For \(d^5\), wider sextet scans repair a four-of-252-root coverage
deficit, supplying 20--21 roots above the fixed 0.5-ligand-hole
threshold with following-root checks. That improves coverage while
changing the represented approximation. In XMS, window changes also
change the state-averaged Fock operator. Neither a complete boundary nor
a small selected-septet increment establishes a converged window.

\section{S5. Bare-anion symmetry and implementation
controls}\label{s5.-bare-anion-symmetry-and-implementation-controls}

\subsection{S5.1 Original OpenMolcas 24.02
control}\label{s5.1-original-openmolcas-24.02-control}

The {[}IrF\(_6\){]}\(^-\) test has no embedding or frozen-ion boundary.
Ir lies at the origin; F coordinates in Å are:

{\def\LTcaptype{none} % do not increment counter
\begin{longtable}[]{@{}lrrr@{}}
\toprule\noalign{}
Centre & \(x\) & \(y\) & \(z\) \\
\midrule\noalign{}
\endhead
\bottomrule\noalign{}
\endlastfoot
1 & 0.89157970 & 1.24093212 & 1.08599220 \\
2 & −1.52046859 & 0.15166461 & 1.08599220 \\
3 & 0.62888889 & −1.39259673 & 1.08599220 \\
4 & 1.52046859 & −0.15166461 & −1.08599220 \\
5 & −0.62888889 & 1.39259673 & −1.08599220 \\
6 & −0.89157970 & −1.24093212 & −1.08599220 \\
\end{longtable}
}

Table S4. Cage coordinates with \(S_6\) symmetry to the displayed
precision. The calculation uses Ir ANO-RCC-VDZP, F ANO-RCC-VDZ, DKH2,
RICD threshold \(10^{-6}\) and charge −1.

The initial equal-weight CASSCF average has four active electrons in
five orbitals, 64 inactive orbitals and the three lowest triplets.
Native RASSCF keywords are \texttt{Thrs\ =\ 1.0d-10\ 1.0d-6\ 1.0d-6} and
\texttt{CIRoot\ =\ 3\ 3\ 1}. The same resulting orbital file is reused
for CI-only three- and twelve-triplet calculations. Real pair components
are rotated by 0° or 15° with \texttt{ROSTate}; overlaps against the
unrotated state file check the intended two-dimensional rotation. MS and
XMS use \texttt{MULTistate\ =\ all} and \texttt{XMULtistate\ =\ all},
IPEA 0.25 or 0 and imaginary shift 0.1 Hartree. Native PT2 outputs
report 36 frozen orbitals.

These are spin-free triplet tests. Define
\(\delta_{\rm sf}=(E_{E,1}+E_{E,2})/2-E_A\). The projected grouping
follows the known symmetric model. For strongly split native states, a
closest-pair descriptor alone is not a symmetry assignment.

{\def\LTcaptype{none} % do not increment counter
\begin{longtable}[]{@{}
  >{\raggedleft\arraybackslash}p{(\linewidth - 10\tabcolsep) * \real{0.1667}}
  >{\raggedleft\arraybackslash}p{(\linewidth - 10\tabcolsep) * \real{0.1667}}
  >{\raggedright\arraybackslash}p{(\linewidth - 10\tabcolsep) * \real{0.1667}}
  >{\raggedleft\arraybackslash}p{(\linewidth - 10\tabcolsep) * \real{0.1667}}
  >{\raggedleft\arraybackslash}p{(\linewidth - 10\tabcolsep) * \real{0.1667}}
  >{\raggedleft\arraybackslash}p{(\linewidth - 10\tabcolsep) * \real{0.1667}}@{}}
\toprule\noalign{}
\begin{minipage}[b]{\linewidth}\raggedleft
Roots
\end{minipage} & \begin{minipage}[b]{\linewidth}\raggedleft
IPEA
\end{minipage} & \begin{minipage}[b]{\linewidth}\raggedright
Method
\end{minipage} & \begin{minipage}[b]{\linewidth}\raggedleft
Native closest pair, 0°
\end{minipage} & \begin{minipage}[b]{\linewidth}\raggedleft
Projected \(\delta_{\rm sf}\), 0°
\end{minipage} & \begin{minipage}[b]{\linewidth}\raggedleft
Projected \(\delta_{\rm sf}\), 15°
\end{minipage} \\
\midrule\noalign{}
\endhead
\bottomrule\noalign{}
\endlastfoot
3 & 0.25 & MS & 1.776 & −2.501 & −2.293 \\
3 & 0.25 & XMS & 11.979 & −7.213 & −7.213 \\
3 & 0 & MS & 2.978 & −1.975 & −1.973 \\
3 & 0 & XMS & 0.439 & −6.862 & −6.862 \\
12 & 0.25 & MS & 1.741 & −2.728 & −2.208 \\
12 & 0.25 & XMS & 13.545 & −4.541 & −4.541 \\
12 & 0 & MS & 2.970 & −2.008 & −2.004 \\
12 & 0 & XMS & 0.680 & −5.631 & −5.631 \\
\end{longtable}
}

Table S5. Original OpenMolcas 24.02 control; energies in meV and IPEA in
Hartree. Three-to-twelve-root projected XMS changes are +2.671768 and
+1.231369 meV. The largest XMS orientation difference is
\(4.2381\times10^{-8}\) meV.

A nine-root window cuts the pair formed by roots 9 and 10. Its
representation residual is about 0.38, compared with
\(1.3\times10^{-7}\) and \(8.5\times10^{-6}\) for the complete three-
and twelve-root groups. Its projected pair remains split and it is
excluded from the valid comparison. The approximately 0.43 eV change of
absolute perturbative dressing between windows is not a
ground-to-excited-state gap change.

\subsection{S5.2 Matched release and build
checks}\label{s5.2-matched-release-and-build-checks}

The fresh version comparison contains eight parameter combinations per
binary: three or twelve triplet roots, IPEA 0 or 0.25 Hartree, and
reference rotations 0° or 15°. Both use the exact original cage,
113-function ANO-RCC basis, point-nucleus DKH2 Hamiltonian, fixed
orbital file, 36 frozen orbitals, RICD threshold \(10^{-6}\) and
imaginary shift 0.1 Hartree. Every reference and perturbative module
must finish normally; native rotation overlaps, Hamiltonian spectra,
basis hashes and absolute energies are checked before comparison.

The MN5 site binary reports version 24.02 with tag 440-g175b3a12d. Its
banner hash resolves to public commit
175b3a12d757e9bd048cad8676a4d198999f0186, but this mapping does not
establish the absence of site patches. It is labeled \textbf{site
24.02+440}, separately from the earlier archived GNU calculation whose
exact bare-anion source tag was not recovered. Stable \textbf{26.10} is
built from clean source commit 32a60dfd712fb951e66e775fd4a515b5b65a5ac3,
with the exact dependency pin and five passing CTest checks. Both fresh
cluster builds use the documented Intel runtime environment. Native
executable and runtime linkage hashes are retained; no software binaries
are bundled.

{\def\LTcaptype{none} % do not increment counter
\begin{longtable}[]{@{}rrlrrrr@{}}
\toprule\noalign{}
Roots & IPEA & Method & Site 0° & Site 15° & 26.10 0° & 26.10 15° \\
\midrule\noalign{}
\endhead
\bottomrule\noalign{}
\endlastfoot
3 & 0.25 & MS & -2.497825 & -2.294745 & -2.500002 & -2.293249 \\
3 & 0.25 & XMS & -7.213716 & -7.213716 & -7.213036 & -7.213036 \\
3 & 0 & MS & -1.975161 & -1.973448 & -1.974889 & -1.973176 \\
3 & 0 & XMS & -6.862100 & -6.862100 & -6.862100 & -6.862100 \\
12 & 0.25 & MS & -2.734519 & -2.205510 & -2.735882 & -2.205105 \\
12 & 0.25 & XMS & -4.541402 & -4.541402 & -4.541540 & -4.541540 \\
12 & 0 & MS & -2.008358 & -2.003865 & -2.008493 & -2.003865 \\
12 & 0 & XMS & -5.630595 & -5.630595 & -5.630731 & -5.630595 \\
\end{longtable}
}

Table S5b. Signed diagnostic \(\delta_{\rm sf}\) in meV for the matched
site and pinned-release runs, using the historical least-squares
representation consistently. The grouping is assigned from the spatial
rotation characters, not from a native closest-pair distance. Group
averaging is a diagnostic transformation.

The maximum difference of native relative levels over every selected
root is 0.003265 meV for MS and 0.004626 meV for XMS. The corresponding
largest differences of projected \(\delta_{\rm sf}\) are 0.002177 and
0.000680 meV. These are observed build-comparison residuals, not
uncertainty estimates or evidence that source changes alone caused them.
Native XMS orientation equality is exact at printed precision for the
three-root cases; small twelve-root differences remain at the
printed/numerical scale. The original archived outputs in Table S5 are
preserved rather than replaced by the fresh values.

\begin{figure}
\centering
\pandocbounded{\includegraphics[keepaspectratio,alt={Matched software-release controls. (a) Signed projected XMS pair-minus-singleton separation for three- and twelve-root windows at both IPEA values, zero reference rotation; the site and pinned-release points nearly overlap. Lines connect specified calculations and do not extrapolate convergence. (b) Maximum absolute difference between native relative levels of stable 26.10 and the site snapshot, over all selected roots, in microelectronvolts. Each spectrum is referenced to its own lowest retained triplet. Tick labels specify root count, IPEA and rotation. These are the sixteen Frozen36 calculations, separate from the Frozen0 independent-implementation protocol. Build and compiler effects are included in the observed difference; no statistical error bars are implied.}]{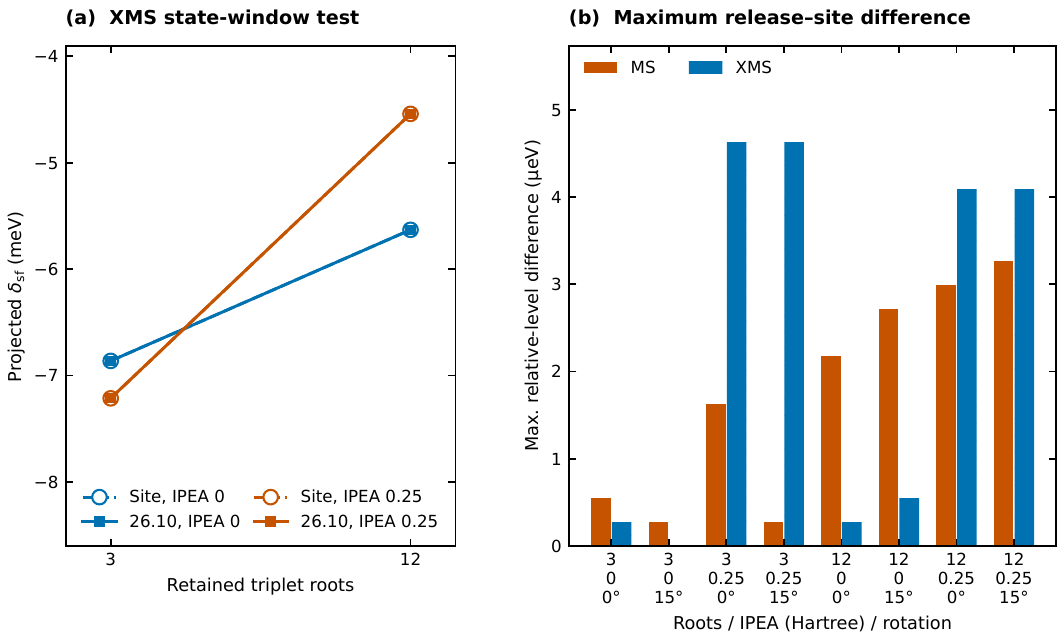}}
\caption{Matched software-release controls. (a) Signed projected XMS
pair-minus-singleton separation for three- and twelve-root windows at
both IPEA values, zero reference rotation; the site and pinned-release
points nearly overlap. Lines connect specified calculations and do not
extrapolate convergence. (b) Maximum absolute difference between native
relative levels of stable 26.10 and the site snapshot, over all selected
roots, in microelectronvolts. Each spectrum is referenced to its own
lowest retained triplet. Tick labels specify root count, IPEA and
rotation. These are the sixteen Frozen36 calculations, separate from the
Frozen0 independent-implementation protocol. Build and compiler effects
are included in the observed difference; no statistical error bars are
implied.}
\end{figure}

Two additional three-root, IPEA-0.25 cases use the official development
binary v26.10-164-g3bb7bcd61 (source commit
3bb7bcd615e9c96a7226803b9695415b802fc106; GNU 4.8.5 compatibility
runtime). One local GNU 15.2.0 build of exact stable 26.10 supplies an
additional zero-rotation compiler check. These are separate cohorts,
excluded from the sixteen-case release comparison. Their executable,
library and source/basis provenance accompanies the native records.

{\def\LTcaptype{none} % do not increment counter
\begin{longtable}[]{@{}
  >{\raggedright\arraybackslash}p{(\linewidth - 10\tabcolsep) * \real{0.1667}}
  >{\raggedleft\arraybackslash}p{(\linewidth - 10\tabcolsep) * \real{0.1667}}
  >{\raggedright\arraybackslash}p{(\linewidth - 10\tabcolsep) * \real{0.1667}}
  >{\raggedleft\arraybackslash}p{(\linewidth - 10\tabcolsep) * \real{0.1667}}
  >{\raggedleft\arraybackslash}p{(\linewidth - 10\tabcolsep) * \real{0.1667}}
  >{\raggedleft\arraybackslash}p{(\linewidth - 10\tabcolsep) * \real{0.1667}}@{}}
\toprule\noalign{}
\begin{minipage}[b]{\linewidth}\raggedright
Additional build
\end{minipage} & \begin{minipage}[b]{\linewidth}\raggedleft
Rotation
\end{minipage} & \begin{minipage}[b]{\linewidth}\raggedright
Method
\end{minipage} & \begin{minipage}[b]{\linewidth}\raggedleft
Second level
\end{minipage} & \begin{minipage}[b]{\linewidth}\raggedleft
Third level
\end{minipage} & \begin{minipage}[b]{\linewidth}\raggedleft
Projected \(\delta_{\rm sf}\)
\end{minipage} \\
\midrule\noalign{}
\endhead
\bottomrule\noalign{}
\endlastfoot
Development & 0° & MS & 5.348398 & 7.123397 & -2.507214 \\
Development & 0° & XMS & 16.564115 & 28.538758 & -7.213037 \\
Development & 15° & MS & 5.302139 & 7.183534 & -2.288624 \\
Development & 15° & XMS & 16.564115 & 28.538758 & -7.213037 \\
Local stable & 0° & MS & 5.356834 & 7.134009 & -2.492927 \\
Local stable & 0° & XMS & 16.554591 & 28.535220 & -7.213308 \\
\end{longtable}
}

Table S5c. Additional build controls; energies in meV relative to the
lowest retained triplet, with IPEA in Hartree as specified above.
Development is not the exact stable 26.10 release. Absolute energy
offsets are supplied in the source records and are not interpreted as
ground-to-excited-state gaps.

A property-only RASSI check adds ONEL to avoid diagonalizing the
deliberately duplicated reference-plus-rotated diagnostic basis. For one
matched three-root case, all original-state angular-momentum matrices
and downstream MS/XMS absolute energies, spectra and effective
Hamiltonians are identical at printed precision. The spurious
redundant-basis RASSI eigenvalues are excluded from the analysis. This
one-case check does not replace the physical RASSI/SOC spectra in the
eight-material campaign.

\subsection{S5.3 Independent implementation and contraction
controls}\label{s5.3-independent-implementation-and-contraction-controls}

The independent calculations use the installed BAGEL 1.2.2 module,
without source modifications by us, and its fully internally contracted
XMS implementation\textsuperscript{9--11}. All use the same cage, point
nuclei, DKH2, Ir ANO-RCC-VDZP/F ANO-RCC-VDZ orbital basis (113 AOs),
charge −1, 132 electrons and triplet CAS(4,5) with 64 closed orbitals.
The original orbital coefficients are restored at native precision from
the Molcas RasOrb. The original reference uses the first 36 frozen
orbitals, but BAGEL diagonalizes the full closed block before selecting
that core; the largest frozen--unfrozen cross-overlap elements are
\(1.47\times10^{-6}\) for three roots and 0.00586 for twelve. Therefore
the independent correlated controls use \textbf{no frozen orbitals in
either code}, and IPEA zero. They do not replace Table S5 or reproduce
its frozen-core approximation.

A supported fixed-subspace input path uses \texttt{algorithm:second},
\texttt{maxiter:1}, \texttt{thresh:1e6} and \texttt{canonical:false},
with \texttt{thresh\_fci:1e-12}, \texttt{maxiter\_fci:200},
\texttt{davidson\_subspace:40} and \texttt{print\_thresh:1e-12}. Finite
initial gradients below the deliberately large break threshold, absence
of interspace orbital microiterations and preserved 64/5/44 projectors
establish its scope; the program's convergence message does not
establish orbital stationarity. Within-block rotations are followed by
fresh CI and density matrices. The first CASSCF stage is a discarded
nonrelativistic initialization; the second stage uses the physical DKH2
reference. Its saved Reference supplies XMS directly. Exported active
occupation fields are not interpreted as physical electron counts: their
source routine writes square-root density metadata. All 50 native
determinants per root instead give four active electrons and \(S^2=2\);
the retained CI/RDMs supply the correlated calculation. Exact inputs,
native outputs, the source audit and the separately preserved
failed-route reproducer are in
\texttt{01\_calculation\_records/implementation\_validation/bagel/}.

The fitting basis spans all primitive-product exponent ranges without
discarding small orbital contraction coefficients. Even-tempered ratios
1.6, 1.4 and 1.2 give 2852, 3987 and 7204 auxiliary functions.
Tightening 1.4 to 1.2 changes the three-root high gap by 0.000272 meV;
the largest twelve-root change is 0.05333 meV, so the correlated test is
limited to three roots. BAGEL removes 2584 near-null directions of the
7204-function metric at threshold \(10^{-12}\); the extremal negative
metric value is \(-9.95\times10^{-12}\), within that near-null region.
The three-root BAGEL gap is 4.605527 meV. Molcas RICD thresholds
\(10^{-6}\), \(10^{-8}\) and \(10^{-10}\) give 4.613418, 4.612602 and
4.611786 meV. The remaining 0.006259 meV cross-code difference
\textbf{fails} the initial 0.001 meV matching target. It is retained as
a numerical limitation, not absorbed into a fitted offset. Native
absolute energies and constants/build provenance accompany the gap data.

{\def\LTcaptype{none} % do not increment counter
\begin{longtable}[]{@{}
  >{\raggedright\arraybackslash}p{(\linewidth - 6\tabcolsep) * \real{0.2500}}
  >{\raggedleft\arraybackslash}p{(\linewidth - 6\tabcolsep) * \real{0.2500}}
  >{\raggedleft\arraybackslash}p{(\linewidth - 6\tabcolsep) * \real{0.2500}}
  >{\raggedleft\arraybackslash}p{(\linewidth - 6\tabcolsep) * \real{0.2500}}@{}}
\toprule\noalign{}
\begin{minipage}[b]{\linewidth}\raggedright
Program and treatment
\end{minipage} & \begin{minipage}[b]{\linewidth}\raggedleft
Imaginary shift (Hartree)
\end{minipage} & \begin{minipage}[b]{\linewidth}\raggedleft
Native lower-pair spacing
\end{minipage} & \begin{minipage}[b]{\linewidth}\raggedleft
Third level above lowest
\end{minipage} \\
\midrule\noalign{}
\endhead
\bottomrule\noalign{}
\endlastfoot
OpenMolcas site MS & 0 & 2.958150 & 7.851301 \\
OpenMolcas site XMS & 0 & 0.453342 & 7.249658 \\
OpenMolcas site MS & 0.1 & 2.853658 & 7.652114 \\
OpenMolcas site XMS & 0.1 & 0.443273 & 7.239317 \\
BAGEL SS--SR XMS & 0 & 0.458936 & 7.247478 \\
BAGEL SS--SR XMS & 0.1 & 0.449105 & 7.237353 \\
BAGEL MS--MR XMS & 0 & 0.000046 & 6.895705 \\
\end{longtable}
}

Table S5d. Distinct three-root, Frozen0, IPEA0 controls; energies in
meV. OpenMolcas uses RICD \(10^{-10}\) and BAGEL the ratio-1.2
auxiliary. Finite-shift conventions differ: BAGEL applies its shift
correction to diagonal effective-Hamiltonian elements. The zero-shift
comparison removes this difference. BAGEL's non-XMS shared-Fock branch
is not a conventional state-specific MS replication and is not included.
No arbitrary external 0°/15° reference rotation is claimed in BAGEL.

The spatial \(C_3\) operation is built from actual AOs and molecular
orbitals, then lifted to all 50 CI determinants with the exact
closed-shell determinant factor. Direct full occupied-minor checks
verify the Schur-complement construction. In the Gram-normalized
three-state CI basis, the original representation has unitarity and cube
residuals of \(1.73\times10^{-9}\) and \(2.54\times10^{-9}\). Its
state-window closure residual is \(2.16\times10^{-6}\) in Frobenius
norm; the active orbital space itself has leakage \(1.27\times10^{-5}\).
These are finite defects, not proof of exact closure. The normalized
real \(A\) vector \(a\) defines \(P_A=aa^T\) and \(P_E=I-P_A\); \(B_E\)
is an orthonormal basis for the two-dimensional \(E\) range. For the
effective Hamiltonian in the same CI gauge we report

\begin{equation}s_E=\lambda_{\max}(B_E^\dagger H B_E)-\lambda_{\min}(B_E^\dagger H B_E),\qquad \kappa=\|[H,D(C_3)]\|_F,\end{equation}

\begin{equation}\eta=\|P_AHP_E\|_F,\qquad \delta_{EA}=\tfrac12\operatorname{Tr}(P_EH)-\operatorname{Tr}(P_AH).\end{equation}

Here \(s_E\) resolves the symmetry-defined block rather than the nearest
native eigenvalues, \(\kappa\) measures covariance, and \(\eta\) retains
mixing with the singleton. The Frobenius norms \(\kappa\) and \(\eta\)
and the \(E\)-sector eigenvalue span are invariant under a common
unitary change of CI gauge; a maximum matrix element is gauge dependent.
The native eigenvalues remain unprojected results. Finite-order repair
of the diagnostic representation changes the projected Hamiltonian by
less than \(8\times10^{-9}\) meV, without removing leakage outside the
represented state space.

{\def\LTcaptype{none} % do not increment counter
\begin{longtable}[]{@{}
  >{\raggedright\arraybackslash}p{(\linewidth - 10\tabcolsep) * \real{0.1667}}
  >{\raggedleft\arraybackslash}p{(\linewidth - 10\tabcolsep) * \real{0.1667}}
  >{\raggedleft\arraybackslash}p{(\linewidth - 10\tabcolsep) * \real{0.1667}}
  >{\raggedleft\arraybackslash}p{(\linewidth - 10\tabcolsep) * \real{0.1667}}
  >{\raggedleft\arraybackslash}p{(\linewidth - 10\tabcolsep) * \real{0.1667}}
  >{\raggedleft\arraybackslash}p{(\linewidth - 10\tabcolsep) * \real{0.1667}}@{}}
\toprule\noalign{}
\begin{minipage}[b]{\linewidth}\raggedright
Reference / contraction
\end{minipage} & \begin{minipage}[b]{\linewidth}\raggedleft
Shift
\end{minipage} & \begin{minipage}[b]{\linewidth}\raggedleft
\(s_E\)
\end{minipage} & \begin{minipage}[b]{\linewidth}\raggedleft
\(\kappa\)
\end{minipage} & \begin{minipage}[b]{\linewidth}\raggedleft
\(\eta\)
\end{minipage} & \begin{minipage}[b]{\linewidth}\raggedleft
\(\delta_{EA}\)
\end{minipage} \\
\midrule\noalign{}
\endhead
\bottomrule\noalign{}
\endlastfoot
Original / SS--SR & 0 & 0.448330 & 0.885376 & 0.283546 & −7.001320 \\
Original / SS--SR & 0.1 & 0.438885 & 0.867611 & 0.278035 & −6.996732 \\
Original / SS--SR, tighter solve & 0.1 & 0.438885 & 0.867607 & 0.278033
& −6.996732 \\
Original / MS--MR & 0 & 0.000047 & 0.000401 & 0.000162 & −6.895680 \\
Minimally C3-closed / SS--SR & 0.1 & 0.464402 & 0.895377 & 0.282311 &
−6.997375 \\
\end{longtable}
}

Table S5e. BAGEL scalar symmetry diagnostics in meV; shift in Hartree.
The original CASCI commutator norm is 0.000644 meV and its \(E\)-block
splitting is 0.000272 meV. The minimally C3-closed CASCI reference gives
0.000182 meV and a splitting unresolved at native eight-decimal energy
precision. The final PT2 energies print ten decimals (a 0.00000272 meV
step); digits below that scale are not independent accuracy claims. No
statistical uncertainties are assigned.

All correlated arms use an orthogonal first-order basis, no frozen
virtuals, \texttt{maxiter:150}, \texttt{maxtile:10},
\texttt{davidson\_subspace:6}, \texttt{thresh\_overlap:1e-9} and
residual threshold \(10^{-10}\), except the explicitly tighter
\(10^{-12}\) arm. All three states stop at iteration 8, or 10 for the
tighter arm. Source inspection establishes the threshold-controlled
early-exit branch; rounded zero residuals in native stdout are not
treated as exact zeros. Complete effective-Hamiltonian spectra match
printed final energies within \(9.4\times10^{-11}\) Hartree. Decimal
subtraction of the large common energy precedes floating-point matrix
analysis. Tightening the solver changes the centered Hamiltonian by at
most \(2.72\times10^{-6}\) meV and leaves final printed energies
identical.

The contraction switch can be stated through its contracted first-order
spans before linear-dependence removal and
orthogonalization\textsuperscript{11--13}:

\begin{equation}
\mathcal V_I^{\mathrm{SS-SR}}=\operatorname{span}_{\mu}\{\hat E_\mu|\widetilde\Phi_I\rangle\},\qquad
\mathcal V_I^{\mathrm{MS-MR}}=\operatorname{span}_{\mu,J}\{\hat E_\mu|\widetilde\Phi_J\rangle\}.
\end{equation}

Here \(\widetilde\Phi_J\) are the XMS-rotated roots, \(J\) runs over the
same three retained references, and \(\mu\) labels individual external
excitation operators and their orbital indices. MS--MR includes
excitations built on every retained root for each target \(I\); SS--SR
retains only root \(I\). This changes the contracted response space
without enlarging the reference window or establishing exact physical
symmetry.

The MS--MR control changes only \texttt{sssr:true} to
\texttt{sssr:false} at zero shift. Its saved Reference, auxiliary basis,
Fock matrix and XMS rotation are identical to the SS--SR arm. The
reduction of \(s_E\) and \(\kappa\) therefore isolates contraction
dependence in this model. A separate minimal orbital-projector repair
reduces physical active-space leakage to \(1.72\times10^{-9}\) and the
fresh three-state closure residual to \(6.01\times10^{-8}\). The largest
active principal-angle change is \(7.36\times10^{-6}\) radians, with
geometry and orbital basis unchanged. Fresh CASCI and its own \(A/E\)
sectors are used before SS--SR dressing. The remaining 0.474961 meV
native spacing shows that the original orbital leakage is not required
for a large defect in this test. It does not show that leakage has no
influence: exact occupied-determinant alignment finds a
\(1.67\times10^{-8}\) Hartree change of the Fock block, a reduction of
its lowest eigenvalue gap from \(2.87\times10^{-8}\) to
\(3.13\times10^{-9}\) Hartree, and about 0.703° change of the
corresponding eigenvector directions. The same-reference MS--MR
comparison avoids that coupled reference change.

These controls narrow attribution to the tested first-order contraction
rather than establish a universal failure of XMS or a newly identified
code error. They do not demonstrate convergence of the embedded material
windows, identify an optimal IPEA value, or replace the native
baseline32 campaign.

\section{S6. Spectral estimators and term
assignments}\label{s6.-spectral-estimators-and-term-assignments}

The fluoride targets are fitted features of 300 K powder RIXS at 35 meV
resolution: K\(_2\)IrF\(_6\), 802 ± 0.3 and 914 ± 1.0 meV;
Cs\(_2\)IrF\(_6\), 804 ± 0.5 and 911 ± 1.5 meV\textsuperscript{14}.
Their quoted fit uncertainties do not represent a tolerance on the
electronic-structure approximation. Vibronic effects can split spectral
features without a static loss of cubic symmetry\textsuperscript{15}.

The executed forward Hamiltonian uses \(q_\nu=b_\nu+b_\nu^\dagger\) and
omits a common oscillator zero-point constant:

\begin{equation}H_{\rm vib}=H_{\rm stat}+\sum_{\nu=1}^{5}\hbar\omega_\nu b_\nu^\dagger b_\nu+g_e(q_1E_\theta+q_2E_\epsilon)+g_t(q_3T_\xi+q_4T_\eta+q_5T_\zeta).\end{equation}

In the \(j=3/2\) basis ordered by \(m=3/2,1/2,-1/2,-3/2\),
\(E_\theta=J_z^2-5I/4\), \(E_\epsilon=(J_x^2-J_y^2)/\sqrt3\), and
\((T_\xi,T_\eta,T_\zeta)=(\{J_y,J_z\},\{J_z,J_x\},\{J_x,J_y\})/\sqrt3\).
The static term is
\(H_{\rm stat}=\delta(T_\xi+T_\eta+T_\zeta)/(2\sqrt3)\), with \(\delta\)
the upper-minus-lower electronic doublet separation; the electronic
midpoint is added to the resulting transition energies. Quadratic
couplings and curvature changes are zero in this comparison.

{\def\LTcaptype{none} % do not increment counter
\begin{longtable}[]{@{}
  >{\raggedright\arraybackslash}p{(\linewidth - 8\tabcolsep) * \real{0.2000}}
  >{\raggedleft\arraybackslash}p{(\linewidth - 8\tabcolsep) * \real{0.2000}}
  >{\raggedleft\arraybackslash}p{(\linewidth - 8\tabcolsep) * \real{0.2000}}
  >{\raggedleft\arraybackslash}p{(\linewidth - 8\tabcolsep) * \real{0.2000}}
  >{\raggedleft\arraybackslash}p{(\linewidth - 8\tabcolsep) * \real{0.2000}}@{}}
\toprule\noalign{}
\begin{minipage}[b]{\linewidth}\raggedright
Cage / coupling sector
\end{minipage} & \begin{minipage}[b]{\linewidth}\raggedleft
\(\hbar\omega_e\)
\end{minipage} & \begin{minipage}[b]{\linewidth}\raggedleft
\(\hbar\omega_t\)
\end{minipage} & \begin{minipage}[b]{\linewidth}\raggedleft
\(g_e\)
\end{minipage} & \begin{minipage}[b]{\linewidth}\raggedleft
\(g_t\)
\end{minipage} \\
\midrule\noalign{}
\endhead
\bottomrule\noalign{}
\endlastfoot
K / CAS & 74.301598 & 30.664811 & 26.420073 & 7.277026 \\
K / 2h0p & 75.397576 & 30.353251 & 25.675210 & 7.686785 \\
Cs / CAS & 72.279005 & 29.595646 & 26.845446 & 7.102955 \\
Cs / 2h0p & 73.428314 & 29.272572 & 26.174181 & 7.494847 \\
\end{longtable}
}

Table S6a. All entries are in meV under the stated \(q=b+b^\dagger\)
normalization. The two \(e_g\) and three \(t_{2g}\) modes share their
sector frequencies and couplings. These parameters originate from
symmetric displacements of the CASSCF-orbital cages. CAS and ORCA DCD
electronic levels use the CAS coupling set; ligand-active MS/XMS and
integer-hole PBE0 2h0p/2h2p levels use the 2h0p set. Couplings are
therefore transferred between specified electronic protocols, not
independently recomputed for each method. Full-precision inputs and the
unchanged matrix convention are supplied in
\texttt{01\_calculation\_records/forward/parameters.json} and its
solver.

Equal weights over four \(\Gamma_8\) electronic parents and harmonic
thermal initial populations determine intensities from overlaps with
coupled final states. The final basis has at most seven total oscillator
quanta; initial populations are checked from two to six. Initial states
with normalized population below \(10^{-6}\) are skipped. A 35 meV FWHM
Gaussian models resolution. Intermediate-state and polarization factors
are absent. The two-component Pearson VII fit uses one shared shape
exponent:

\begin{equation}I(E)=\sum_{j=A,B}A_j\left[1+\left(\frac{E-E_j}{w_j}\right)^2(2^{1/m}-1)\right]^{-m}.\end{equation}

Here \(w_j\) is the half-width at half-maximum. The grid extends from
\(-300\) to \(450\) meV relative to the electronic midpoint in 0.25 meV
steps. The two strongest local maxima exceeding 15\% of the envelope
maximum, ordered as \(p_A<p_B\), define a fit interval
\([p_A-120,p_B+120]\) meV. The envelope is normalized to its global
maximum. Bounds are \(0\le A_j\le2\), \(p_j-60\le E_j\le p_j+60\) meV,
\(2\le w_j\le120\) meV and \(1\le m\le50\). Initial centres are
\(p_A,p_B\), widths are 20/25 meV, amplitudes are the normalized
envelope at the maxima, and \(m=2\). Nonlinear least squares uses at
most 40,000 function evaluations; fixed \(m=1,2,50\) fits test shape
sensitivity. The fitted comparison uses the positive static sign; the
underlying forward records also contain both signs. Exact scripts, input
records and stored checkpoints accompany the SI.

{\def\LTcaptype{none} % do not increment counter
\begin{longtable}[]{@{}lrll@{}}
\toprule\noalign{}
Cage / model & Static split & Envelope A / B & Fitted A / B \\
\midrule\noalign{}
\endhead
\bottomrule\noalign{}
\endlastfoot
K / CASSCF CAS & 2.7 & 968.1 / 1059.1 & 968.2 / 1059.0 \\
K / CASSCF 2h0p & 0.1 & 939.2 / 1031.2 & 939.4 / 1031.0 \\
K / Ligand MS & 66.3 & 733.9 / 849.4 & 744.3 / 849.7 \\
K / Ligand XMS & 46.8 & 746.2 / 848.7 & 752.4 / 849.1 \\
K / PBE0 CAS & 47.1 & 783.9 / 885.9 & 789.5 / 886.1 \\
K / PBE0 2h0p & 10.7 & 870.2 / 962.2 & 870.5 / 962.2 \\
K / PBE0 2h2p & 1.1 & 918.6 / 1010.6 & 918.9 / 1010.4 \\
K / ORCA DCD & 34.8 & 943.7 / 1038.5 & 945.2 / 1038.5 \\
Cs / CASSCF CAS & 6.9 & 972.6 / 1061.9 & 972.7 / 1062.2 \\
Cs / CASSCF 2h0p & 4.9 & 944.5 / 1034.7 & 944.6 / 1034.9 \\
Cs / Ligand MS & 34.5 & 774.0 / 868.2 & 775.5 / 868.3 \\
Cs / Ligand XMS & 36.9 & 762.7 / 857.7 & 764.6 / 857.9 \\
Cs / PBE0 CAS & 44.9 & 789.9 / 888.9 & 793.9 / 889.2 \\
Cs / PBE0 2h0p & 11.9 & 875.3 / 966.0 & 875.5 / 966.0 \\
Cs / PBE0 2h2p & 6.3 & 924.3 / 1014.6 & 924.3 / 1014.7 \\
Cs / ORCA DCD & 20.0 & 950.5 / 1040.5 & 950.6 / 1040.9 \\
\end{longtable}
}

Table S6b. Distinct spectral estimators in meV for all sixteen
material/model cases, with six-quanta initial populations and a shared
free Pearson VII exponent. Adjacent models do not form one convergence
sequence. None puts both fitted centres within the chosen ±20 meV
comparison interval; this statement concerns the specified
transferred-coupling forward model.

\begin{figure}
\centering
\pandocbounded{\includegraphics[keepaspectratio,alt={Absolute spectral estimators for all sixteen fluoride material/model cases. Grey ticks show static doublet energies, blue circles the two envelope maxima with a six-quanta initial cutoff, and open orange diamonds the Pearson VII fitted centres. Green bands show the measured 300 K fitted features and their quoted fit uncertainties. Each row retains its own electronic model; adjacent rows do not form a convergence sequence. The forward operator omits intermediate-state and polarization factors.}]{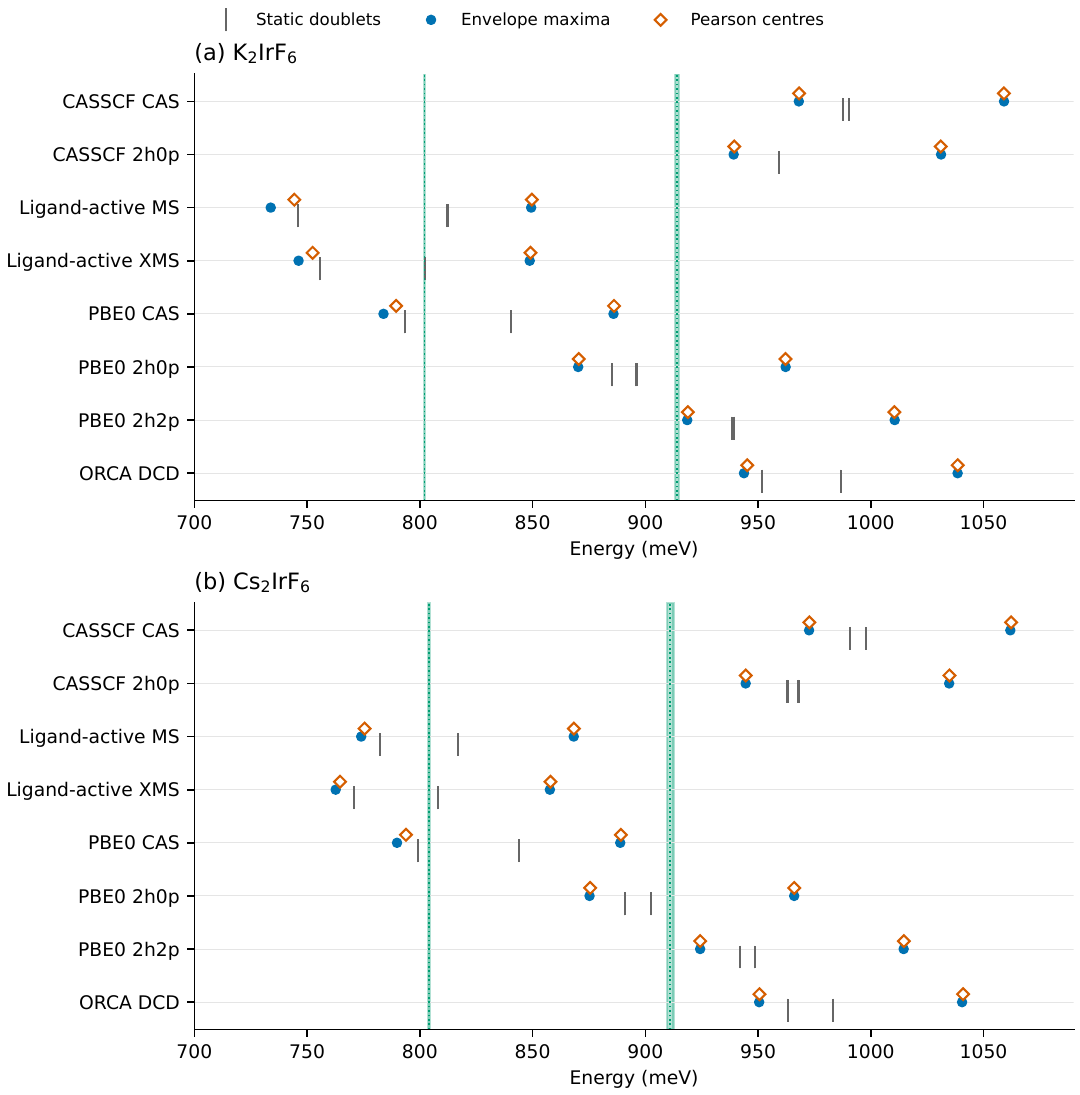}}
\caption{Absolute spectral estimators for all sixteen fluoride
material/model cases. Grey ticks show static doublet energies, blue
circles the two envelope maxima with a six-quanta initial cutoff, and
open orange diamonds the Pearson VII fitted centres. Green bands show
the measured 300 K fitted features and their quoted fit uncertainties.
Each row retains its own electronic model; adjacent rows do not form a
convergence sequence. The forward operator omits intermediate-state and
polarization factors.}
\end{figure}

\FloatBarrier

The six-quanta initial cutoff retains 99.19--99.42\% of the harmonic
thermal weight and moves direct maxima by at most 0.5 meV relative to
the smaller cutoff. For K ligand-active MS, the lower envelope maximum
is 733.9 meV and fitted centre 744.3 meV. The fit reaches the tested
shape-exponent bound \(m=50\); changing estimator alters that centre by
3.5 meV. Thermal convergence and estimator stability are separate
checks. Values formed by adding a model-dependent vibronic offset to a
measured maximum are not independent electronic targets.

For osmates, native term weights assign states before comparison with
optical origins\textsuperscript{16}. The K\(_2\)OsCl\(_6\) metal-shell
XMS \(^1T_2\) centroid is 1522.152743 meV with mean weight 0.728203; its
\(^1A_1\) state is 2224.095797 meV with weight 0.873843. Residuals
against 1311/2117 meV are +16.106235\%/+5.058847\%. Structural phase and
temperature must be retained: a measured low-temperature noncubic
splitting is not itself evidence of numerical symmetry breaking.

The two osmate optical-origin comparisons are retained together to
distinguish the material trend from the RIXS-maximum comparison:

{\def\LTcaptype{none} % do not increment counter
\begin{longtable}[]{@{}
  >{\raggedright\arraybackslash}p{(\linewidth - 6\tabcolsep) * \real{0.2500}}
  >{\raggedleft\arraybackslash}p{(\linewidth - 6\tabcolsep) * \real{0.2500}}
  >{\raggedright\arraybackslash}p{(\linewidth - 6\tabcolsep) * \real{0.2500}}
  >{\raggedright\arraybackslash}p{(\linewidth - 6\tabcolsep) * \real{0.2500}}@{}}
\toprule\noalign{}
\begin{minipage}[b]{\linewidth}\raggedright
Compound
\end{minipage} & \begin{minipage}[b]{\linewidth}\raggedleft
Optical origin
\end{minipage} & \begin{minipage}[b]{\linewidth}\raggedright
Ligand MS / residual
\end{minipage} & \begin{minipage}[b]{\linewidth}\raggedright
Ligand XMS / residual
\end{minipage} \\
\midrule\noalign{}
\endhead
\bottomrule\noalign{}
\endlastfoot
K\(_2\)OsCl\(_6\) & 350 & 350.818 / +0.818 & 341.469 / −8.531 \\
K\(_2\)OsBr\(_6\) & 349 & 347.644 / −1.356 & 341.443 / −7.557 \\
\end{longtable}
}

Table S6c. Energies and calculated-minus-optical residuals in meV.
Origins are the experimental values in Table I of
Ref.\textsuperscript{16}, distinct from the 341/340 meV RIXS maxima. The
chloride origin uses the 50 K cubic phase; low-temperature phase-induced
splittings must not be treated as numerical defects of the cubic model.
The source supplies no uncertainty for these extracted origin values, so
the residuals are not uncertainty-normalized accuracy scores.

\section{S7. Complete finite-space
response}\label{s7.-complete-finite-space-response}

The variational Hamiltonian holds orbital integrals fixed while changing
the determinant sector. With chemists' two-electron integrals,

\begin{equation}H=E_{\rm core}+\sum_{pq\sigma}h_{pq}a^\dagger_{p\sigma}a_{q\sigma}+\frac12\sum_{pqrs\sigma\tau}(pq|rs)a^\dagger_{p\sigma}a^\dagger_{r\tau}a_{s\tau}a_{q\sigma}+H_{\rm SO}.\end{equation}

Following a restricted-sector construction\textsuperscript{17}, the
selected 2h0p sector permits at most two holes in five filled ligand
orbitals and no occupation of the external radial group. Complete
dimensions are 12,180 for \(d^4\) and 7,752 for \(d^5\). Each
Hamiltonian uses its own moment operator,
\(m_\alpha=\mu_\alpha/\mu_B=-(L_\alpha+g_eS_\alpha)\), with
\(g_e=2.00231930436\).

For energies in joules and \(p_n=e^{-\beta E_n}/Z\), the static molar SI
response follows the Kubo expression\textsuperscript{18},

\begin{equation}\chi_{\alpha\beta}=\mu_0N_A\mu_B^2\left[\sum_{nm}K_{nm}\operatorname{Re}(m^\alpha_{nm}m^\beta_{mn})-\beta\langle m_\alpha\rangle\langle m_\beta\rangle\right],\end{equation}

\begin{equation}K_{nm}=\begin{cases}(p_n-p_m)/(E_m-E_n),&E_n\ne E_m,\\ \beta p_n,&E_n=E_m.\end{cases}\end{equation}

The analytic degenerate limit avoids division by artificial tiny
splittings. Powder response is \(\operatorname{Tr}\chi/3\). For a
low-state projector \(P\) and \(Q=1-P\), complementary solves
\(Q(H-E_n)Q|x_{n\beta}\rangle=Qm_\beta|n\rangle\) supply higher virtual
transitions. Their contribution
\(2\sum_{n\in P}p_n\operatorname{Re}\langle n|m_\alpha Q|x_{n\beta}\rangle\)
is added once to the explicit low-state sum. Omitted thermal weight
requires a separate bound. Finite fields test the convention
\(H(B)=H(0)-\boldsymbol\mu\cdot\mathbf B\).

\Needspace{18\baselineskip}

{\def\LTcaptype{none} % do not increment counter
\begin{longtable}[]{@{}lrrr@{}}
\toprule\noalign{}
Material & CAS & 2h0p & Increase (\%) \\
\midrule\noalign{}
\endhead
\bottomrule\noalign{}
\endlastfoot
LiIrF\(_6\) & 665.354 & 710.297 & 6.755 \\
Ba\(_2\)YIrO\(_6\) & 760.962 & 868.565 & 14.140 \\
K\(_2\)OsCl\(_6\) & 868.003 & 1040.170 & 19.835 \\
K\(_2\)OsBr\(_6\) & 869.805 & 1117.146 & 28.436 \\
K\(_2\)IrF\(_6\) & 1274.227 & 1457.386 & 14.374 \\
Cs\(_2\)IrF\(_6\) & 1264.593 & 1453.319 & 14.924 \\
K\(_2\)IrCl\(_6\) & 1184.858 & 1365.149 & 15.216 \\
K\(_2\)IrBr\(_6\) & 1204.674 & 1389.554 & 15.347 \\
\end{longtable}
}

Table S7. Local electronic powder susceptibilities at 300 K in
\(10^{-6}\) cm\(^3\) mol\(^{-1}\). Multiply a cgs value expressed in
cm\(^3\) mol\(^{-1}\) by \(4\pi\times10^{-6}\) to obtain m\(^3\)
mol\(^{-1}\). The family spectral consistency residual is
\(9.03\times10^{-6}\) meV; the largest calculated thermal-omission bound
is \(2.45\times10^{-9}\) in the displayed response units.

The temperature dependence provides a separate check from the
single-temperature comparison. The \(d^4\) curves remain nearly constant
on this interval, while the \(d^5\) local Kramers models have a strong
low-temperature response. These calculations contain no intersite
ordering or fitted background; their low-temperature curves are not a
prediction of a bulk paramagnetic phase.

\begin{figure}
\centering
\pandocbounded{\includegraphics[keepaspectratio,alt={Local electronic powder susceptibility of all eight compounds. Colours identify materials within each panel; dashed and solid curves denote metal-shell CAS and complete 2h0p on the same orbitals within a material. The d\^{}5 ordinate is logarithmic. The points underlying the curves are the archived calculated temperatures; lines only join them. Exchange, magnetic ordering, defects and bulk background contributions are absent.}]{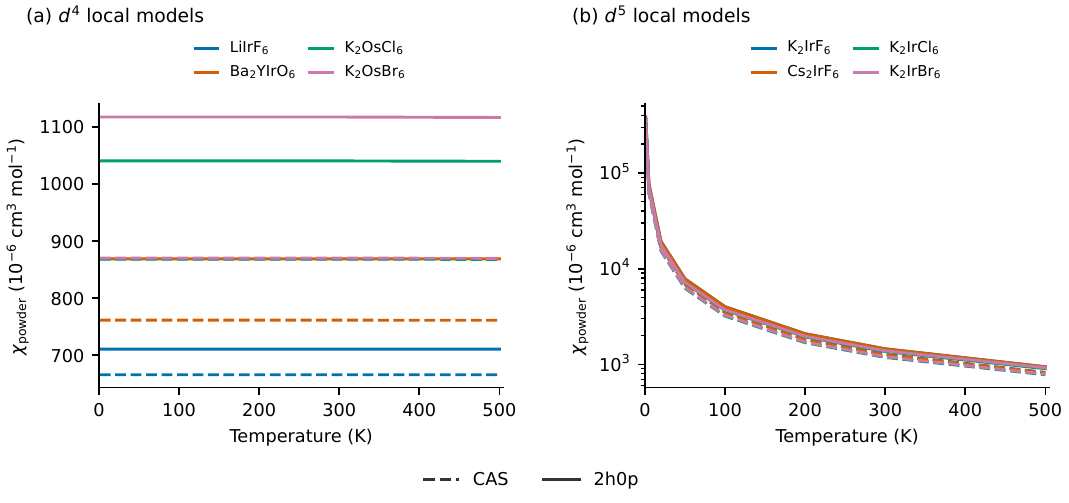}}
\caption{Local electronic powder susceptibility of all eight compounds.
Colours identify materials within each panel; dashed and solid curves
denote metal-shell CAS and complete 2h0p on the same orbitals within a
material. The \(d^5\) ordinate is logarithmic. The points underlying the
curves are the archived calculated temperatures; lines only join them.
Exchange, magnetic ordering, defects and bulk background contributions
are absent.}
\end{figure}

\FloatBarrier

The two Li response families must remain separate: family A uses the
earlier 132-electron, coreless-cation CASSCF-derived VDZP boundary,
whereas family B uses the 144-electron PBE0-derived reference with six
explicit Li cores and the enlarged boundary. In the family PBE0 orbital
reference the centroid rises from 437.510 to 478.019 meV while
susceptibility increases as shown above. In a corresponding-orbital
reference the centroid falls from 548.140 to 532.146 meV and
low-temperature powder response rises from 541.90 to 637.34. Their
basis/boundary choices differ, so these pairs are not successive steps
toward a common limit. A distinct matched orbital switch controls basis,
geometry and embedding and still reverses the sign of the centroid
increment. Similar ligand-plus-metal union subspaces do not guarantee
identical restricted sectors.

\Needspace{17\baselineskip}

The matched orbital switch can be read directly from its endpoints:

{\def\LTcaptype{none} % do not increment counter
\begin{longtable}[]{@{}
  >{\raggedright\arraybackslash}p{(\linewidth - 8\tabcolsep) * \real{0.2000}}
  >{\raggedleft\arraybackslash}p{(\linewidth - 8\tabcolsep) * \real{0.2000}}
  >{\raggedleft\arraybackslash}p{(\linewidth - 8\tabcolsep) * \real{0.2000}}
  >{\raggedleft\arraybackslash}p{(\linewidth - 8\tabcolsep) * \real{0.2000}}
  >{\raggedleft\arraybackslash}p{(\linewidth - 8\tabcolsep) * \real{0.2000}}@{}}
\toprule\noalign{}
\begin{minipage}[b]{\linewidth}\raggedright
Orbital reference
\end{minipage} & \begin{minipage}[b]{\linewidth}\raggedleft
CAS centroid
\end{minipage} & \begin{minipage}[b]{\linewidth}\raggedleft
2h0p centroid
\end{minipage} & \begin{minipage}[b]{\linewidth}\raggedleft
2h0p minus CAS
\end{minipage} & \begin{minipage}[b]{\linewidth}\raggedleft
2h0p pair-minus-singleton
\end{minipage} \\
\midrule\noalign{}
\endhead
\bottomrule\noalign{}
\endlastfoot
CASSCF & 550.629669 & 534.488886 & -16.140784 & 12.286878 \\
Integer-hole PBE0 & 437.510305 & 478.018547 & +40.508242 & 12.018313 \\
\end{longtable}
}

Table S7b. Controlled Li orbital-reference comparison; all entries are
in meV. Geometry, the 305-function VDZP orbital basis, embedding,
electron number and configuration-sector rule are fixed between rows;
the selected orbital subspaces change. Within each row, CAS and 2h0p
instead retain the same orbitals and integrals while the allowed
configurations change. The last column is the signed
pair-minus-singleton energy descriptor; individual levels retain their
finite pair defects. Digits identify deterministic model results, not
physical error bars.

The two 2h0p centroids differ by 56.470338 meV, whereas their
pair-minus-singleton descriptors differ by 0.268565 meV. AO-metric
principal angles, obtained from the singular values of
\(C_A^\dagger S C_B\), reach 8.131° for the five metal-like orbitals,
7.782° for the five selected occupied ligand orbitals and only 2.406°
for their ten-orbital union. Thus a similar union can have a different
internal partition, and the same nominal restricted-sector rule acts on
different many-electron spaces. This identifies the changed reference
without uniquely apportioning its energy effect.

\begin{figure}
\centering
\pandocbounded{\includegraphics[keepaspectratio,alt={Matched orbital-reference intervention. First centroids and signed pair-minus-singleton separations compare CASSCF and integer-hole PBE0 orbitals at fixed geometry, 305-function basis, electron count and embedding. The right panel shows all AO-metric principal angles for metal-like, selected occupied-ligand and combined subspaces. The similar union does not imply an identical restricted many-electron space. This controlled switch is separate from the four response models below.}]{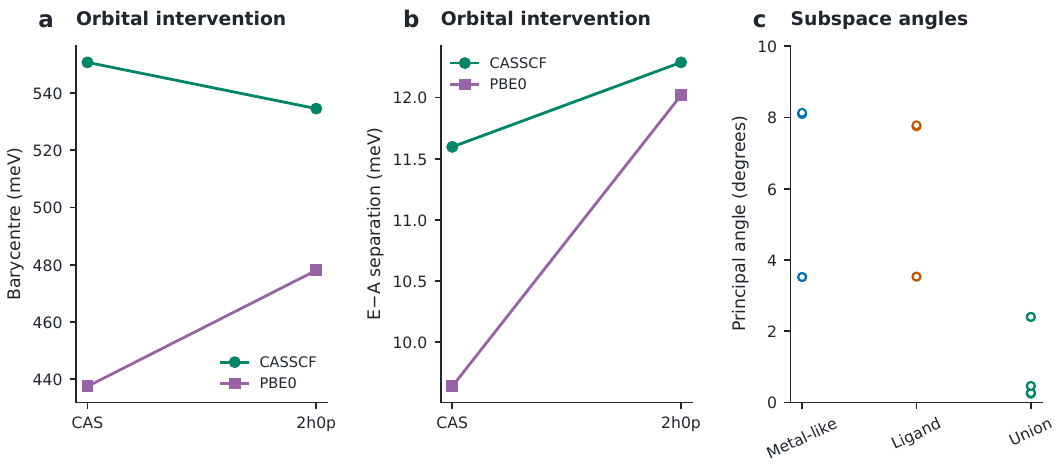}}
\caption{Matched orbital-reference intervention. First centroids and
signed pair-minus-singleton separations compare CASSCF and integer-hole
PBE0 orbitals at fixed geometry, 305-function basis, electron count and
embedding. The right panel shows all AO-metric principal angles for
metal-like, selected occupied-ligand and combined subspaces. The similar
union does not imply an identical restricted many-electron space. This
controlled switch is separate from the four response models below.}
\end{figure}

\FloatBarrier

\Needspace{22\baselineskip}

For context, the retained configuration and dressing paths are:

{\def\LTcaptype{none} % do not increment counter
\begin{longtable}[]{@{}
  >{\raggedright\arraybackslash}p{(\linewidth - 2\tabcolsep) * \real{0.5000}}
  >{\raggedright\arraybackslash}p{(\linewidth - 2\tabcolsep) * \real{0.5000}}@{}}
\toprule\noalign{}
\begin{minipage}[b]{\linewidth}\raggedright
Intervention / reference
\end{minipage} & \begin{minipage}[b]{\linewidth}\raggedright
First-centroid path (meV)
\end{minipage} \\
\midrule\noalign{}
\endhead
\bottomrule\noalign{}
\endlastfoot
Configurations, corresponding-orbital response family & CAS 548.140 →
0h2p 542.648 → 1h1p 546.273 → 2h0p 532.146 → 2h2p 523.814 \\
Configurations, integer-hole PBE0 family & CAS 437.510 → 1h0p 500.104 →
2h0p 478.019 → 1h1p 550.600 → 2h1p 526.925 → 2h2p 522.486 \\
Perturbative dressing, ligand-active reference & CASCI 479.537 → SS
538.946 → MS 460.591 → XMS 458.073 \\
\end{longtable}
}

Table S7c. Distinct Li interventions, rounded to 0.001 meV. Each
configuration row fixes its own orbitals, integrals, geometry, basis and
embedding. In \(h\mathrm h p\mathrm p\), at most \(h\) holes are
permitted in the five occupied ligand orbitals and at most \(p\)
particles in the five external radial orbitals; the metal group remains
unrestricted at the fixed total electron number. The
corresponding-orbital response family in the first row has a different
boundary from the matched CASSCF row in Table S7b. The PBE0 row uses the
same reference as Table S7b. The dressing row retains its ligand-active
reference while changing the correlated treatment; SS denotes
state-specific CASPT2, with IPEA 0.25 and imaginary shift 0.1 Hartree.
These paths separate configuration, orbital and perturbative changes;
neither cross-row subtraction nor their listed order supplies a
convergence or accuracy estimate.

\Needspace{17\baselineskip}

The four-model response comparison also resolves the anisotropy and
virtual-state contributions at 1 K:

{\def\LTcaptype{none} % do not increment counter
\begin{longtable}[]{@{}
  >{\raggedright\arraybackslash}p{(\linewidth - 8\tabcolsep) * \real{0.2000}}
  >{\raggedleft\arraybackslash}p{(\linewidth - 8\tabcolsep) * \real{0.2000}}
  >{\raggedleft\arraybackslash}p{(\linewidth - 8\tabcolsep) * \real{0.2000}}
  >{\raggedleft\arraybackslash}p{(\linewidth - 8\tabcolsep) * \real{0.2000}}
  >{\raggedleft\arraybackslash}p{(\linewidth - 8\tabcolsep) * \real{0.2000}}@{}}
\toprule\noalign{}
\begin{minipage}[b]{\linewidth}\raggedright
Family / sector
\end{minipage} & \begin{minipage}[b]{\linewidth}\raggedleft
First centroid (meV)
\end{minipage} & \begin{minipage}[b]{\linewidth}\raggedleft
\(\chi_\parallel\)
\end{minipage} & \begin{minipage}[b]{\linewidth}\raggedleft
\(\chi_\perp\)
\end{minipage} & \begin{minipage}[b]{\linewidth}\raggedleft
\(\chi_{\rm powder}\)
\end{minipage} \\
\midrule\noalign{}
\endhead
\bottomrule\noalign{}
\endlastfoot
A / CAS & 548.140 & 553.484 & 536.115 & 541.905 \\
A / 2h0p & 532.146 & 651.098 & 630.454 & 637.335 \\
B / CAS & 437.510 & 686.022 & 655.020 & 665.354 \\
B / 2h0p & 478.019 & 735.160 & 697.865 & 710.297 \\
\end{longtable}
}

Table S7d. Susceptibilities are in \(10^{-6}\) cm\(^3\) mol\(^{-1}\);
\(\chi_\perp=(\chi_{xx}+\chi_{yy})/2\) and \(\chi_\parallel=\chi_{zz}\).
A and B are the 132- and 144-electron response families specified above,
not the two rows of the strictly matched switch in Table S7b. Each
CAS/2h0p pair fixes its own reference. Family B retains nominal
transverse-pair defects of 0.149 and 0.083 meV, respectively.

For A/2h0p, the first excited trio contributes 620.8437 and 599.9194 to
\(\chi_\parallel\) and \(\chi_\perp\), respectively; higher virtual
states contribute 30.2542 and 30.5348. The higher-state powder share is
about 4.8\% although thermal occupation is negligible at 1 K. This is
why reproducing the first energy manifold does not by itself reproduce
the response.

\begin{figure}
\centering
\pandocbounded{\includegraphics[keepaspectratio,alt={Four Li finite-model responses at 1 K. Panels show opposite centroid shifts, increased powder response in both families, axial anisotropy and the first-trio/higher-virtual-state decomposition. Open and filled markers denote CAS and 2h0p. Reference A is the earlier 132-electron CASSCF-derived model; B is the 144-electron PBE0-derived model. Cross-family differences include orbital, basis and boundary changes. They do not define a pure orbital intervention or a convergence path. The plotted values are unchanged from the archived calculations.}]{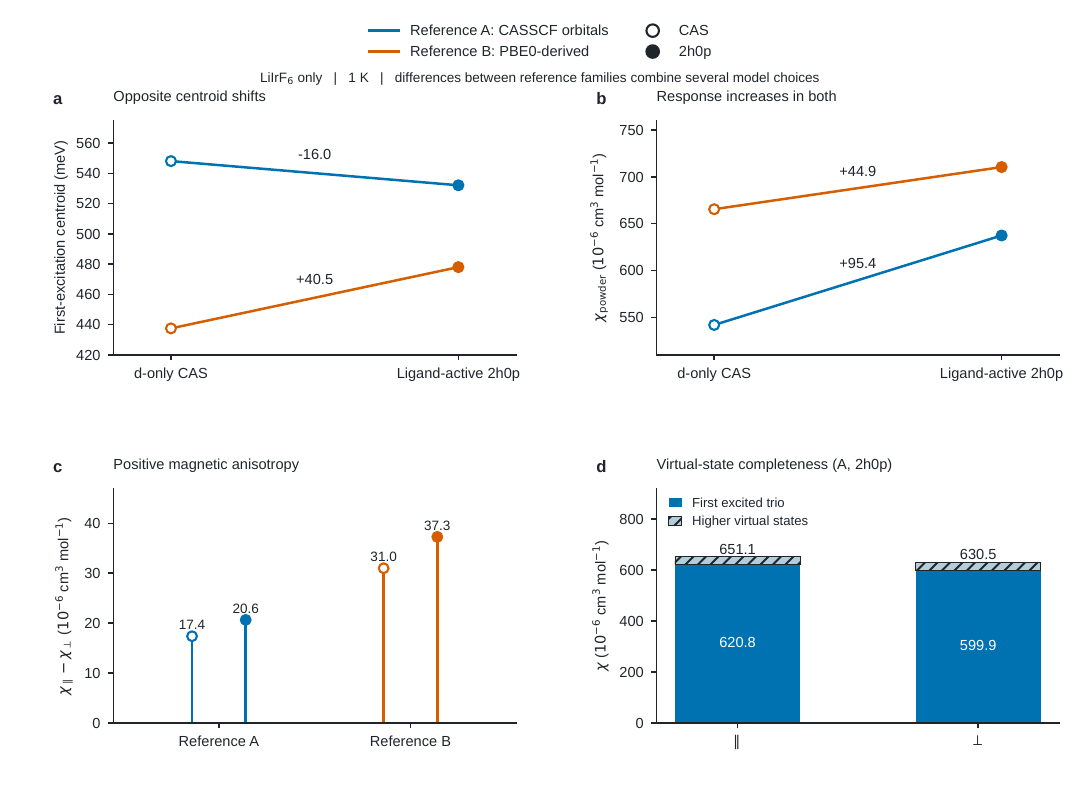}}
\caption{Four Li finite-model responses at 1 K. Panels show opposite
centroid shifts, increased powder response in both families, axial
anisotropy and the first-trio/higher-virtual-state decomposition. Open
and filled markers denote CAS and 2h0p. Reference A is the earlier
132-electron CASSCF-derived model; B is the 144-electron PBE0-derived
model. Cross-family differences include orbital, basis and boundary
changes. They do not define a pure orbital intervention or a convergence
path. The plotted values are unchanged from the archived calculations.}
\end{figure}

\FloatBarrier

\subsection{S7.1. Finite fields and thermal
completeness}\label{s7.1.-finite-fields-and-thermal-completeness}

For field-dependent levels \(E_n(B)\) in joules, the single-cluster
partition function, free energy and moment satisfy

\begin{equation}\begin{aligned}Z(B,T)&=\sum_n e^{-\beta E_n(B)},\qquad F=-k_BT\ln Z,\\ M_\alpha&=-\frac{\partial F}{\partial B_\alpha}=k_BT\frac{\partial\ln Z}{\partial B_\alpha}.\end{aligned}\end{equation}

At zero field the molar electronic heat capacity is

\begin{equation}C_V^{\rm el}=\frac{N_A}{k_BT^2}\left(\langle E^2\rangle-\langle E\rangle^2\right).\end{equation}

Each SOC component is counted once. All metal-shell CAS states are
retained. In the detailed Li 2h0p calculations, 25 low eigenpairs were
obtained, with the last reserved as a thermal-tail check; an independent
31-root random-start calculation checked the spectrum. Complementary
Sternheimer solves supply higher virtual transitions without a physical
denominator shift. Independent \(\pm0.5\) T calculations check the
zero-field convention, and \(\pm5\) T at 500 K check the thermal
response. Spin and orbital moments are included consistently; orbital
diamagnetic \(B^2\) terms are not.

The Li series evaluates 41 temperatures from 1 to 900 K using stored
spectra and response kernels. At the highest temperature, five
additional 2h0p states enter complementary-response checks; the largest
omitted-response bound is \(3.35\times10^{-12}\) in susceptibility units
\(10^{-6}\) cm\(^3\) mol\(^{-1}\), and omitted partition weight is below
\(10^{-18}\). The four Li models give \(C_V^{\rm el}=1.09\)--2.99 J
mol\(^{-1}\) K\(^{-1}\) at 900 K. These numerical tail bounds apply to
the specified finite electronic models, not to omitted lattice or
intersite physics.

A separate eight-material calculation evaluates complete metal-shell CAS
spectra at the same 41 temperatures. Its overlap with the original
temperature series agrees within \(10^{-6}\) in the displayed
susceptibility units, and field reversal at 10 T gives \(M(+B)=-M(-B)\)
within \(7.2\times10^{-10}\,\mu_B\). This CAS-only calculation is
distinct from the matched CAS/2h0p curves above.

\begin{figure}
\centering
\pandocbounded{\includegraphics[keepaspectratio,alt={Complete metal-shell CAS temperature and field response of eight compounds. Each curve contains 41 directly evaluated temperatures from 1 to 900 K. The left panels show local powder susceptibility; the right panels show the moment along the stored z axis in a 10 T field. Every material uses its own finite Hamiltonian and magnetic operators. No ligand-active 2h0p curve appears in this figure. Low-temperature intersite ordering and high-temperature structural changes are outside the model.}]{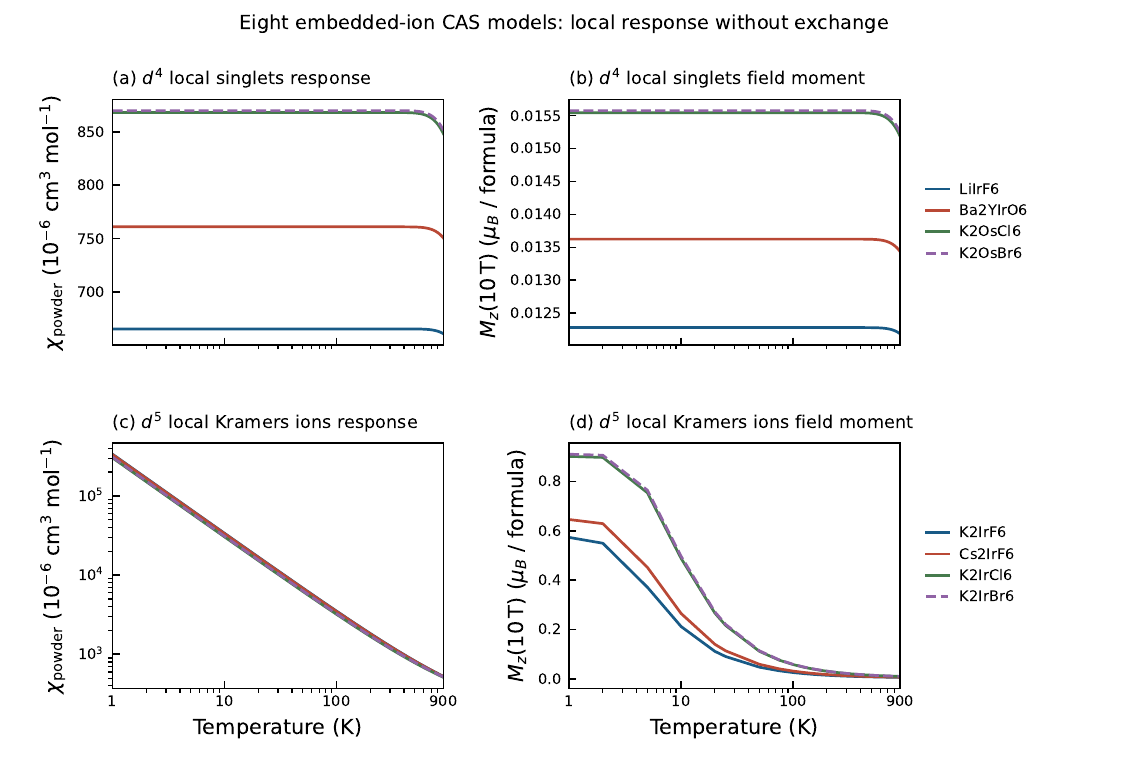}}
\caption{Complete metal-shell CAS temperature and field response of
eight compounds. Each curve contains 41 directly evaluated temperatures
from 1 to 900 K. The left panels show local powder susceptibility; the
right panels show the moment along the stored \(z\) axis in a 10 T
field. Every material uses its own finite Hamiltonian and magnetic
operators. No ligand-active 2h0p curve appears in this figure.
Low-temperature intersite ordering and high-temperature structural
changes are outside the model.}
\end{figure}

\FloatBarrier

\subsection{S7.2. Comparison with extracted electronic
components}\label{s7.2.-comparison-with-extracted-electronic-components}

{\def\LTcaptype{none} % do not increment counter
\begin{longtable}[]{@{}
  >{\raggedright\arraybackslash}p{(\linewidth - 10\tabcolsep) * \real{0.1667}}
  >{\raggedleft\arraybackslash}p{(\linewidth - 10\tabcolsep) * \real{0.1667}}
  >{\raggedleft\arraybackslash}p{(\linewidth - 10\tabcolsep) * \real{0.1667}}
  >{\raggedleft\arraybackslash}p{(\linewidth - 10\tabcolsep) * \real{0.1667}}
  >{\raggedleft\arraybackslash}p{(\linewidth - 10\tabcolsep) * \real{0.1667}}
  >{\raggedleft\arraybackslash}p{(\linewidth - 10\tabcolsep) * \real{0.1667}}@{}}
\toprule\noalign{}
\begin{minipage}[b]{\linewidth}\raggedright
Material
\end{minipage} & \begin{minipage}[b]{\linewidth}\raggedleft
Published component
\end{minipage} & \begin{minipage}[b]{\linewidth}\raggedleft
CAS at 300 K
\end{minipage} & \begin{minipage}[b]{\linewidth}\raggedleft
2h0p at 300 K
\end{minipage} & \begin{minipage}[b]{\linewidth}\raggedleft
CAS residual (\%)
\end{minipage} & \begin{minipage}[b]{\linewidth}\raggedleft
2h0p residual (\%)
\end{minipage} \\
\midrule\noalign{}
\endhead
\bottomrule\noalign{}
\endlastfoot
K\(_2\)OsCl\(_6\) & 960 & 868.003 & 1040.170 & -9.58 & +8.35 \\
K\(_2\)OsBr\(_6\) & 1020 & 869.805 & 1117.146 & -14.72 & +9.52 \\
Ba\(_2\)YIrO\(_6\) & 751 & 760.962 & 868.565 & +1.33 & +15.65 \\
\end{longtable}
}

Table S7e. Susceptibilities are in \(10^{-6}\) cm\(^3\) mol\(^{-1}\);
residuals are \(100(\chi_{\rm calc}/\chi_{\rm ref}-1)\). The osmate
references are rounded Van Vleck plateaux and the Ba reference is the
extracted electronic component for the cited
sample\textsuperscript{16,19}. They are not raw total-susceptibility
measurements exactly at 300 K. Adding ligand holes improves the absolute
component comparison for the two osmates and worsens it for this Ba
reference, so increased response is not a universal accuracy
improvement.

The cited sources do not supply extraction uncertainties for these
rounded components. No background is fitted here. Local electronic
response excludes intersite order, core diamagnetism and impurity terms;
comparisons cannot be generalized to the total measured
susceptibility\textsuperscript{20--23}.

\section{S8. Independent treatments, model reduction and limits of
transfer}\label{s8.-independent-treatments-model-reduction-and-limits-of-transfer}

The common-seed ORCA comparison uses 132 electrons, the same Li
geometry/boundary and identical CASSCF root weights. All 210 CASSCF SOC
levels agree between the method preparations. The scalar reference
energy is −18405.776125447 Hartree and orbital gradient
\(1.485\times10^{-4}\), below the calculation's \(2.5\times10^{-4}\)
threshold.

{\def\LTcaptype{none} % do not increment counter
\begin{longtable}[]{@{}lrr@{}}
\toprule\noalign{}
Treatment & Centroid (meV) & Closest pair (meV) \\
\midrule\noalign{}
\endhead
\bottomrule\noalign{}
\endlastfoot
CASSCF & 545.344084 & 0.001240 \\
SC-NEVPT2 in QD pilot & 566.094492 & 0.117785 \\
QD-NEVPT2 & 559.703107 & 0.560409 \\
DCD order 0, diagnostic & 531.199218 & 0.000364 \\
DCD order 1 & 559.484228 & 0.000318 \\
\end{longtable}
}

Table S8. Independent treatment comparison\textsuperscript{24--28}.
These are not the embedded OpenMolcas ligand-active reference and do not
supply an independent MS/XMS implementation. The QD result is not
accepted as a symmetry repair merely because its centroid is close to
DCD.

A computed-parent reduction provides a different numerical check.
Splitting a known Hamiltonian into retained \(W\) and complementary
\(R\) blocks gives
\(H_{\rm eff}(E)=H_{WW}+H_{WR}(E-H_{RR})^{-1}H_{RW}\). This identity
holds to about \(10^{-15}\) Hartree for the tested 488-state parent.
Second-order reductions above 10 eV reproduce first/second centroids
within 0.0234/0.0542 meV when the complementary block is diagonalized.
These bounds apply inside that parent, not to configurations absent from
it. Observable reduction also requires transformed magnetic operators.

Full-shell fitting supplies another warning about transferring centroids
to fine structure. Table S8b compares the native DCD1 spectrum with two
reconstructions using its fitted scalar crystal field and Slater
interaction: spherical SOC, or the printed native anisotropic SOC
matrices. The independent 210-determinant replay matches all rounded
native spherical-model utility levels within 2.9 micro-eV, while the
global spin-free fitting RMS residuals remain 81.6/88.2 meV for Li/Cs.
Thus successful numerical replay and faithful mapping are separate
properties.

\Needspace{19\baselineskip}

{\def\LTcaptype{none} % do not increment counter
\begin{longtable}[]{@{}
  >{\raggedright\arraybackslash}p{(\linewidth - 6\tabcolsep) * \real{0.2500}}
  >{\raggedleft\arraybackslash}p{(\linewidth - 6\tabcolsep) * \real{0.2500}}
  >{\raggedleft\arraybackslash}p{(\linewidth - 6\tabcolsep) * \real{0.2500}}
  >{\raggedleft\arraybackslash}p{(\linewidth - 6\tabcolsep) * \real{0.2500}}@{}}
\toprule\noalign{}
\begin{minipage}[b]{\linewidth}\raggedright
Salt / treatment
\end{minipage} & \begin{minipage}[b]{\linewidth}\raggedleft
First centroid (meV)
\end{minipage} & \begin{minipage}[b]{\linewidth}\raggedleft
Second centroid (meV)
\end{minipage} & \begin{minipage}[b]{\linewidth}\raggedleft
Signed pair-minus-singleton descriptor (meV)
\end{minipage} \\
\midrule\noalign{}
\endhead
\bottomrule\noalign{}
\endlastfoot
Li, native DCD1 & 559.484 & 937.719 & +12.612 \\
Li, fitted scalar + spherical SOC & 566.199 & 918.432 & -61.378 \\
Li, fitted scalar + native SOC & 564.051 & 925.337 & +60.393 \\
Cs, native DCD1 & 559.764 & 938.221 & +10.634 \\
Cs, fitted scalar + spherical SOC & 563.422 & 912.649 & -54.254 \\
Cs, fitted scalar + native SOC & 564.901 & 923.132 & +62.679 \\
\end{longtable}
}

Table S8b. Full-shell AILFT mapping diagnostics for the stated
references, not new material predictions. The six-decimal-Hartree
crystal-field printout bounds excitation/separation rounding errors by
0.545 meV, far below the descriptor changes. Pair-minus-singleton values
describe each row's near-pair grouping; sorted levels alone do not
establish state correspondence. ``Native SOC'' changes the SOC matrix
inside the fitted scalar model and does not restore the exact ab initio
scalar Hamiltonian. These examples diagnose the specified mappings
rather than a universal failure of AILFT\textsuperscript{28}.

Energy-only parameter fitting leaves unconstrained directions. In a
full-shell Li mapping, changes summarized by \(F^2\to1.05F^2\),
\(F^4\to0.96706F^4\) and \(H_{\rm SO}\to0.99375H_{\rm SO}\) alter the
first two centroids by +0.0129/−0.0369 meV while changing a
fine-structure descriptor by +3.735 meV. The opposite change keeps both
centroids within 0.041 meV and changes the descriptor by −3.968 meV.
Full precision is used in the numerical records. The underlying DCD1
scalar mapping has an 81.6 meV RMS residual for Li, so this is a
nonidentifiability example within an imperfect model, not a material
parameter determination.

An analytic sign check is also useful. For two effective \(l=1\) holes
with positive SOC \(\xi\) and \(J_H\geq0\),
\(H=\xi\sum_i\mathbf l_i\cdot\mathbf s_i-2J_H\mathbf S^2-(J_H/2)\mathbf L^2+\mathrm{constant}\)
gives

\begin{equation}\Delta_1=\frac{\sqrt{25J_H^2+10J_H\xi+9\xi^2}-5J_H}{2},\qquad \partial\Delta_1/\partial J_H<0.\end{equation}

At fixed SOC this model's too-large gap cannot be explained by a
too-large Hund coupling. The sign does not determine derivatives of the
ligand-extended cluster.

Reproduction files distinguish numerical-table validation, native matrix
reconstruction and a fresh electronic-structure run. The included
deterministic checker recomputes centroids, spans, fitted-feature
residuals and response changes from copied source records, with source
checksums. Native input/matrix locations and the unchanged exact-model
driver are indexed separately. The matched pinned-release OpenMolcas
grid and independent BAGEL XMS contraction controls are supplied
separately in
\texttt{01\_calculation\_records/implementation\_validation/}. They
extend the bare-anion evidence without certifying all embedded state
windows or producing a full fluoride RIXS cross section or bulk magnetic
calculation. The cross-code reference mismatch, finite closure residuals
and untested material-specific convergence remain explicit limits rather
than additional error bars.

\clearpage

\section{S9. Operator provenance and excitation
energies}\label{s9.-operator-provenance-and-excitation-energies}

All excitation energies below are measured from the ground state of the
same operator: \(\epsilon_i=E_i-E_0\). The first-manifold barycentre is
\(\Delta_1=(\epsilon_A+\epsilon_{E,1}+\epsilon_{E,2})/3\), the signed
separation is \(s=(\epsilon_{E,1}+\epsilon_{E,2})/2-\epsilon_A\), and
the pair defect is \(d_E=|\epsilon_{E,2}-\epsilon_{E,1}|\) when the
state assignments are justified. Thus
\(\Delta_1=(E_A+E_{E,1}+E_{E,2})/3-E_0\) for total eigenvalues; no
further ground-state subtraction is applied to tabulated excitation
energies.

\setcounter{table}{8}
\renewcommand{\thetable}{S\arabic{table}}
\begin{table}[htbp]
\centering
\small
\caption{\textbf{Distinct LiIrF$_6$ operators used in the two studies.} Energies in meV. Values marked imposed are model inputs; the other values are calculated. A closest-pair defect in a raw spectrum is not a symmetry-assigned A--E separation.}
\label{tab:operatorprovenance}
\begin{tabular}{p{0.37\linewidth}rrp{0.28\linewidth}}
\toprule
operator & $\Delta_1$ & $s$ & status and reference\\
\midrule
embedded VQZP 2h0p & 529.337 & 11.522 & principal finite-space cluster\\
isolated VQZP 2h0p & 524.052 & 4.419 & matched isolated cluster\\
native Wannier--Kanamori & 354.671 & 18.276 & local spectrum of the intersite model\\
phenomenological local matching & 514 & 11 & imposed local spectrum\\
ligand-active MS-CASPT2, raw & 460.591 & --- & closest-pair defect 11.990\\
same MS operator, $C_3$-projected & 456.040 & 13.66 & diagnostic at zero reference rotation\\
\bottomrule
\end{tabular}
\end{table}

The isolated and embedded VQZP pairs use the measured Li cage,
DKH2/ANO-RCC integrals and the 45-triplet CASSCF(4,5) orbital average.
The embedded model has 144 explicit electrons including six Li\(^+\)
cores; the isolated anion has 132. The 2h0p sector correlates fourteen
electrons in the five selected occupied ligand combinations and five
metal orbitals, allowing up to two ligand holes. SOC acts in the
complete declared determinant space. Threefold symmetry identifies
states; these rows are not obtained by projecting a perturbative
spectrum. The embedding, orbital selection and full spaces are specified
in the crystal-control study.

The separate Wannier--Kanamori model uses \(U=3\) eV, \(J_H=0.4\) eV and
eight electrons in twelve spin orbitals for its two-site parent (495
determinants). Its local SOC and crystal-field terms come from the
Wannier Hamiltonian; it has neither the cluster boundary nor its
multireference correlation treatment. The imposed 514/11 meV variant
replaces the local spectrum while retaining the projected bond terms.
Neither variant is a correction to the VQZP cluster energies.

The MS rows use the embedded 144-electron Li reference, ANO-RCC-VDZP, a
ligand-active CAS(14,10), IPEA 0.25 and a 0.1 Hartree imaginary shift.
The 33-singlet/32-triplet/15-quintet state window and AMFI SOC treatment
are identical before and after projection. Projection changes the
operator, so the raw centroid must not be paired with the projected
splitting. The root-resolved native inputs and representations are
retained in the multiplet-validation archive. These rows describe
calculated operators, not measured spectra. \FloatBarrier

The cluster and intersite rows are from the separate crystal-control
study\textsuperscript{29}. The MS rows are defined in S1--S4 above.
Archive file identifiers and hashes are listed in
\texttt{05\_provenance/MODEL\_PROVENANCE.json}. This comparison
identifies the operators; it does not assign a common physical error to
them.

\section{Data and code availability}\label{data-and-code-availability}

The numerical records and reproduction code supporting this
Supplementary Information are included in the Paper 2 Zenodo archive,
version 1.0.0, DOI:
\href{https://doi.org/10.5281/zenodo.23917145}{10.5281/zenodo.23917145}.
The numbered folders separate calculation records
(\texttt{01\_calculation\_records/}), validation scripts
(\texttt{02\_validation\_scripts/}), figure and table sources
(\texttt{03\_figures\_and\_tables/}), the manuscript and Supplementary
Information (\texttt{04\_manuscript\_and\_SI/}), and provenance
(\texttt{05\_provenance/}). Public access begins when the deposit is
published.

\section*{References}\label{references}
\addcontentsline{toc}{section}{References}

\protect\phantomsection\label{refs}